\documentclass[dvipsnames,svgnames,x11names]{bmcart}

\usepackage{xcolor}

\usepackage{amsthm,amsmath,amssymb}
\usepackage{amsfonts}
\usepackage[width=0.9\textwidth]{caption}
\usepackage{moreverb,url}

\RequirePackage[authoryear]{natbib}

\usepackage[utf8]{inputenc} 
\usepackage{graphicx} 
\usepackage{subcaption} 

\usepackage{parskip}
\makeatletter
\usepackage{comment}
\let\wfs@comment@comment\comment
\let\comment\@undefined
\usepackage[markup=underlined,final]{changes}
\let\wfs@changes@comment\comment
\let\comment\@undefined

\newcommand\comment{%
    \ifthenelse{\equal{\@currenvir}{comment}}
    {\wfs@comment@comment}
    {\wfs@changes@comment}%
}

\makeatother

\newcommand{\e}{\textrm{e}}
\newcommand{\Oh}{\mathcal{O}}

\newcommand{\note}[2][]{\added[#1,comment={#2}]{}}

\begin{document}
\begin{frontmatter}

\begin{fmbox}
\dochead{Research}

\title{Is academia becoming more localised? The growth of regional knowledge networks within international research collaboration}

\author[
 addressref={OxMI}, 
 noteref={n1}, 
 email={fitzgeraldj@maths.ox.ac.uk} 
]{\inits{JA}\fnm{John} \snm{Fitzgerald}}
\author[
 addressref={OxMI, OxOII, TI},
 email={sanna.ojanpera@oii.ox.ac.uk}
]{\inits{SO}\fnm{Sanna} \snm{Ojanper\"a}}
\author[
 addressref={OxMI, CASA, TI}, 
 email={n.oclery@ucl.ac.uk}
]{\inits{NO}\fnm{Neave} \snm{O'Clery}}

\address[id=OxMI]{
 \orgname{Mathematical Institute, Oxford University}, 
 \street{Andrew Wiles Building, Radcliffe Observatory Quarter}, %
 \postcode{OX2 6GG}, 
 \city{Oxford}, 
 \cny{UK} 
}
\address[id=OxOII]{
 \orgname{Oxford Internet Institute, University of Oxford}, 
 \street{Oxford}, %
 \postcode{ASD},
 \city{Oxford},
 \cny{United Kingdom} 
 }
\address[id=CASA]{
 \orgname{The Bartlett Centre for Advanced Spatial Analysis, University College London}, 
 \street{London}, %
 \cny{United Kingdom} 
}
\address[id=TI]{%
 \orgname{The Alan Turing Institute},
 \city{London},
 \cny{United Kingdom}
}
\begin{artnotes}
\note[id=n1]{Corresponding author} 
\end{artnotes}
\end{fmbox}

\begin{abstractbox}

\begin{abstract} 

It is well-established that the process of learning and capability building is core to economic development and structural transformation. Since knowledge is `sticky', a key component of this process is learning-by-doing, which can be achieved via a variety of mechanisms including international research collaboration. Uncovering significant inter-country research ties using Scopus co-authorship data, we show that within-region collaboration has increased over the past five decades relative to international collaboration. Further supporting this insight, we find that while communities present in the global collaboration network before 2000 were often based on historical geopolitical or colonial lines, in more recent years they increasingly align with a simple partition of countries by regions. These findings are unexpected in light of a presumed continual increase in globalisation, and have significant implications for the design of programmes aimed at promoting international research collaboration and knowledge diffusion.   

\end{abstract}


\end{abstractbox}

\end{frontmatter}

\section{Introduction}

Following advances in transportation and communication technology over the past centuries, we're witnessing a rise in global interactions both in terms of cross-border trade and investment as well as flows of people and information. Termed globalisation, this complex process involves interaction and integration of people, businesses, and governments and while it primarily concerns economic aspects, social and cultural dimensions are similarly salient. In parallel to the process of globalisation, ongoing global economic restructuring has resulted in a transition towards a knowledge-based economy, where a `greater reliance on intellectual capabilities than on physical or natural resources' \citep{powell_knowledge_2004} has meant a rise in production and services that are based on knowledge-intensive activities. The importance of specialised skills along with knowledge and information as forms of non-physical capital has grown, since economic growth increasingly derives from intangible intellectual property including copyrights, patents, trademarks, and trade secrets that work to make more effective use of inputs and available resources. Knowledge diffusion and innovation underpin competition and fuel economic growth at almost all stages of development, as well as play a critical role in enabling responses to complex economic, environmental, and social challenges. Domestic and global research communities are central players in creating and diffusing knowledge and contributing to the development of new products and processes. These communities comprise of research and development activities, research laboratories, universities, and other educational institutions that, together with partners in the private sector and government, form innovation ecosystems \citep{jackson_what_2011}. 

The role of innovation, science and technology as drivers of economic growth and as vital enablers of sustainability is highlighted in the recent \citet{united_nations_educational_scientific_and_cultural_organization_unesco_2015} Science Report, which showcases the trajectories of a large number of countries `incorporating science, technology, and innovation in their national development agendas, in order to be less reliant on raw materials and move towards knowledge economies.' The desirability of fostering local skills and capacities for economic development is similarly echoed by recent work in economic complexity and economic geography \citep{hidalgo_product_2007,frenken2007theoretical, HidalgoHausmann2009, hausmann2011network, oclery_commuting_2019},  analysing the growth of cities and regions. This literature finds the availability of diverse knowledge capacities or complex skills and capabilities as central to the development trajectories of regions, countries, and cities. They conceptualise knowledge and capabilities as geographically `sticky', since tacit knowledge and abilities are a result of a workforce with skills learned on-the-job, and are thus not easily transportable. Research collaboration, in particular with academics from other regions likely in possession of novel or complementary skills and capabilities, could allow countries to upgrade their academic capacity and respond to unique societal and economic challenges more readily.

As countries and regions find themselves at various stages of the transformation towards - and readiness to join - the global knowledge economy \citep{ojanpera_engagement_2017,  ojanpera_digital_2019}, the creation of scientific knowledge is more important than ever. Public and private sector funding is directed towards developing domestic research capabilities, and countries are putting policies in place to attract scientific talent from abroad \citep{united_nations_educational_scientific_and_cultural_organization_unesco_2015}. The OECD Development strategy, implemented in partnership with the United Nations and the World Bank, as well as the OECD policy frameworks for Tertiary Education, Innovation, Development, and Gender Equity, all call for the promotion of regional and international research networks in order to further the dual pursuit of research communities everywhere, summarised by the \citet{programme_on_innovation_higher_education_and_research_research_2012} as: `knowledge generation per se and their specific role in attaining national development priorities'. 

Reflecting this trend, the number of researchers and publications has been growing, with a 20 percent increase between 2007 and 2014 \citep{united_nations_educational_scientific_and_cultural_organization_unesco_2015}. The extent of scientific collaboration has increased in parallel, both overall \citep{wagner-dobler_continuity_2001, meyer_commonalities_2004}, and internationally, between researchers based in different countries \citep{narin_scientific_1991, wagner_mapping_2005, wuchty_increasing_2007, jones_multi-university_2008, gazni_mapping_2012}. Various factors have been suggested as underpinning the growing propensity to collaborate, including advancements in technologies facilitating remote collaboration \citep{ding_impact_2010}, policy initiatives and funding schemes to encourage international collaboration \citep{frenken_death_2009, ubfal_impact_2011}, specialisation requiring collaboration with researchers who may not be available within the local talent pool, cultivation of research impact and credibility \citep{kumar_co-authorship_2015}, and avoidance of duplicating research efforts \citep{katz_what_1997}. 

Indeed, the internationalisation of research collaborations has received increasing attention over the past few decades. Collaboratively authored research has higher impact than research published by a sole author, both in terms of number of publications \citep{katz_what_1997, lee_impact_2005, wuchty_increasing_2007} and citations \citep{sooryamoorthy_types_2017, gazni_investigating_2011}, while research published by international author teams tends to attract more citations than research authored by national teams \citep{narin_scientific_1991, katz_what_1997, frenken_citation_2005}. Furthermore, \citet{jones_multi-university_2008} show that multi-university collaborations produce the highest impact papers when top-tier universities are included, and are increasingly stratified by in-group university rank. 

An emergent body of literature on research collaboration networks - reviewed below - has primarily investigated ties between individuals or institutions, often focusing on particular disciplinary communities or bounded by a regional or sub-national context. Few studies, however, have looked in detail at changing patterns of international collaboration focusing on bilateral ties at the country level, and including all major disciplines. Instead studies tend to focus on particular disciplines, such as medicine or the life sciences. We look at research collaboration across all major disciplines, as it reflects the broad creation and diffusion of knowledge, which contributes to the development of new products and processes, or innovation across economic, social, and political domains.

The existing body of research on international research collaboration networks has deployed a variety of network methods, including network visualisation, local network measures focusing on the importance of nodes, models explaining network growth, and regression methods. In the present study, we apply a range of sophisticated methods deriving from network science and mathematical modelling, including historical profile clustering, calculation of \added{the}\deleted{raw and weighted} entropy of collaborations, community detection, and mutual information comparisons, which allow us to uncover patterns that have previously remained opaque. 

Further, where studies have analysed a time period rather than investigated a snapshot, the time window tends to not span more than a decade or two. We address this research gap through analysing a dataset of international collaboratively authored scientific publications covering a range of disciplines published between 1970 and 2018. In doing so, we assess the extent to which countries learn from each other through `borrowing' capabilities and specialisms from colleagues in other countries or regions, and thus induce knowledge flow. In the analysis to follow, we exploit a variety of network and mathematical modelling tools to analyse the temporal evolution of the global collaboration network to reveal what we term `knowledge basins' (a concept related to `skill basins' as proposed by \citet{oclery_modular_2019}). These are groups of countries which tend to collaborate frequently internally, but less frequently with other groups, thus forming localised (and potentially isolated) clusters of research output. These clusters evolve over time, aligning with colonial and historical geopolitical alliances pre-2000, but coalescing \added{more} along geographical or regional lines since 2000. 

The remainder of this paper is organised as follows. Section~\ref{sec:litrev} will survey the relevant research on co-authorship networks. Our choice of data will be elaborated upon in Section 3, while Section 4 will introduce some preliminary analysis of the data. In Sections 5 and 6 we will present our main research methodology and results with some discussion. Finally, Section 7 summarises our contribution, discusses the implications of our findings, and proposes avenues for future work.

\section{Literature review \label{sec:litrev}}

The literature on research networks has its roots in scientometrics, a sub-field of bibliometrics measuring and analysing scientific literature, but it additionally draws from related disciplines of information systems, information science, and science of science. While the creation of the Science Citation Index in 1964 and related studies \citep{burton_half-life_1960, garfield_new_1963, kessler_analysis_1962, osgood_characteristics_1963, price_little_1963, tukey_keeping_1962} were seminal in establishing the field, the pioneering article by \citet{price_networks_1965} was the first one to investigate networks of scientific papers, and found that the network under study was scale-free with the in-degree (citations within an article) and out-degree (citations to an article) having power-law distributions. Since these early studies' focus on citation networks, the literature has branched out to comprise research on varied themes such as co-citation networks (documents are connected if they appear together in a reference list), co-word networks (words are connected if they appear together within a document), research collaboration (in particular through co-authorship of documents or collaborative grants), researcher mobility, and institutional boundaries. While these studies investigate varied topics, some themes that have received substantial research attention include identifying research fronts, evaluating the impact of individual authors in comparison to collaborations, and the relative influence of disciplines and journals. 

\subsection{Knowledge flows and co-authorship networks} 

This paper contributes to the literature on research collaboration - and specifically co-authorship - networks. In many cases, these are thought to be a proxy for knowledge flows, which are inherently challenging to define and measure. 
By knowledge we mean the creation and retention of knowledge by individuals or organisations, and by knowledge flows we mean the exchange or diffusion of ideas by individuals or organizations \citep{jaffe_international_1998}. Such `pure' knowledge and knowledge flows tend to be disembodied, and are non-rivalrous in the sense that one's consumption of knowledge does not prevent another from consuming the same knowledge. While these kinds of knowledge are difficult to measure chiefly due to their disembodied nature, some have suggested that the flows of certain knowledge-intensive products such as citations to patents could work as `windows' to knowledge flows \citep{jaffe_international_1998}. In a similar vein, internationally co-authored publications, which are considered a reliable proxy for research collaboration \citep{melin_studying_1996, glanzel_analysing_2005, heinze_across_2008}, may be considered as `windows' into knowledge flows between researchers located in different countries.

Co-authorship networks are some of the largest publicly available social networks and while they have received somewhat less research attention than citation networks, they enable a close examination of key aspects of what \citet{newman_coauthorship_2004} terms as `the structure of both academic knowledge and academic society'. The existing literature on co-authorship networks can roughly be divided into three streams based on the methodological approaches utilised, namely, bibliometric methods, survey-based methods, and network analysis. The studies applying a network analysis methodology form a somewhat more recent research area, and as our study falls within this stream, we will focus our discussion on the literature using related methodologies.

This literature investigates networks that vary in size from small groups \emph{e.g.} related to a research institution \citep{fagan_assessing_2018} to massive graphs \emph{e.g.} depicting international patent citation networks \citep{de_rassenfosse_sources_2020}. The research field has gained notable interest after three seminal articles from \citet{newman_scientific_2001, newman_scientific_2001-1, newman_structure_2001}, which studied the micro and macro characteristics of seven large scientific co-authorship networks, and an article by \citet{barabasi_evolution_2002} which examined the evolution and dynamics of these networks. Among further studies which looked at researcher collaboration networks, many focused on detecting popular or well positioned individuals \citep{fatt_structure_2010, racherla_social_2010, ye_cross-institutional_2012, santos_co-authorship_2016}. \citet{newman_scientific_2001-1} noted that scientific networks are highly clustered, with many triangles, while \citet{goh_betweenness_2003} found that authors with a high betweenness centrality avoid collaboration with other authors who are similarly well-positioned, and rather seek less connected individuals.  

Focusing on classifying the network structure, \citet{newman_structure_2001} demonstrated that co-authorship networks could be characterised by the `small world' property \emph{i.e.,} each author is not more than five or six steps away from each other within the network. \citet{goh_classification_2002} found that the node degree distribution is scale free, indicating that while most authors have few collaborations, there are some that have numerous collaborations. Finding a similar pattern, \citet{newman_coauthorship_2004} noted that biological scientists have significantly more coauthors than those publishing in mathematics or physics. Various studies have looked at the existence and size of the `giant component', which seems to vary significantly across disciplines. \citet{newman_structure_2001} found it comprises over 90 percent of authors in biomedical research, while \citet{yan_applying_2009} found it comprises just 20 percent of authors in library and information sciences. \citet{hou_structure_2008} studied the network of authors within scientometrics and found that the two largest research clusters work on the same topic, but utilise different methodological approaches. Comparing network communities to the socioeconomic characteristics of the scholars, \citet{rodriguez_relationship_2008} found that communities best align with individuals working in the same department or institution suggesting that co-authorship is primarily driven by departmental and institutional affiliation.

\subsubsection{International research collaboration} 

Studies adopting an international comparison include both regionally and globally focused approaches. Investigating the growth of international collaboration, \citet{wagner_network_2005} argue that the principle of preferential attachment---where those with more collaborations keep attracting proportionally more new collaborations---explains the phenomenon. In support of this hypothesis, \citet{ribeiro_growth_2018} identify a scale free node degree distribution for a global collaboration network comprising various scientific disciplines. Some authors argue that the core leading group consisting of the United States and Western nations has widened to include a much larger number of countries during the 1990's and 2000's \citep{leydesdorff_international_2013}. Other studies focusing on international research collaborations find that geographical distance and national borders continue to hinder cross-border collaboration \citep{frenken_death_2009, doria_arrieta_quantifying_2017}. Looking at the patterns of medical research in Latin America and the Caribbean, \citet{chinchilla-rodriguez_international_2012} find that the most productive countries collaborate mainly internally or with neighbouring countries, while small or developing countries tend to collaborate more distantly. Other studies suggest that the globalisation of science does not seem to have evolved uniformly across all countries and regions, as historical, sociotechnological, and geographical factors continue to play a key role \citep{geuna_global_2015, scherngell_geography_2013}. This existing body of research adopts either a temporal snapshot into global research collaboration or covers a time window spanning up to two decades.

\subsubsection{Data sources} 

Previous research has made use of bibliographic databases, academic search engines (ASEs), and services that offer a combination of these two functions. Bibliographic databases are comprehensive and reliable collections of information on academic outputs which allow users to efficiently query for information. ASEs on the other hand use computer algorithms to search the internet and recognize items which correspond to a query. They are less structured and subject to inconsistencies yet tend to be significantly larger in scope.
 
While it is challenging to measure the reach of these datasets, a recent article by \citet{gusenbauer_google_2019} attempted to measure their respective sizes. The two largest scholarly bibliographic databases include Scopus (72m records) and Web of Science (67m records). The ASEs offer some significantly larger datasets, and comprise, among others, Google's academic index Google Scholar (387m records), WorldWideScience (323m records), AMiner (232m records), Microsoft Academic (171m records), Bielefeld Academic Search Engine (BASE) (118m records), Q-Sensei Scholar (55m records), and Semantic Scholar (40m records). Aggregate services include ProQuest (280m records) and EbscoHost (132m records). While these sources of data have gained popularity within the field \citep{harzing_google_2016}, each has their advantages and limitations depending on the geographic, disciplinary, or temporal scale of interest.

\section{The Scopus database}

Our dataset contains all co-authorship relations between authors of documents published between 1970 and 2018 which are indexed in Scopus. We chose Scopus as our data source because it has a high level of accuracy as is characteristic for bibliographic databases \citep{gusenbauer_google_2019, gusenbauer_which_2019}. It also has wide geographic, disciplinary, and temporal coverage including 24,600 active titles and 5,000 publishers of scientific journals, books, and conference proceedings across the fields of science, technology, medicine, social sciences, and arts and humanities \citep{elsevier_scopus_2020}. Since we sought as comprehensive a dataset as possible, we decided not to consider the academic search engines because, while they are able to access the largest number of records, the query functions for them seem to be unreliable for detailed bibliometric data such as author affiliation \citep{mingers_normalizing_2017, gusenbauer_google_2019}. Similarly, while the aggregate services ProQuest and EbscoHost and the bibliographic database Web of Science provide more accurate results, it was not apparent whether our institutional access to these services would cover all constituent databases (a well-known shortcoming of these services \citep{gusenbauer_google_2019}). 

While there are obvious advantages to using the Scopus dataset, there are nonetheless several known limitations including weaker coverage for the social sciences and humanities, and non-English publications \citep{aksnes_criteria-based_2019}. 
However, some have argued \citep{bennett_english_2013} that English has come to dominate academia as a `lingua franca' leading to erosion of scholarly discourses in other languages and possibly introducing preferences for certain kinds of knowledge \citep{trahar_hovering_2019}. 
While quantifying this trend isn't possible within the scope of this analysis, it is likely to introduce a shift in original contributions from other languages to English over time and thus might increase the representativeness of our data. Furthermore, while Scopus does not include all possible academic outputs, the categories indexed are arguably some of the most salient kinds of academic outputs, and we would not expect that other omitted categories of outputs would introduce a specific geographic bias into our findings. 

Since we are interested in collaborative relationships on a country level and across scientific disciplines, we first produce a dataset including all publications with authors in multiple countries (including papers with authors affiliated to multiple institutions in different countries), and aggregate this data to form yearly counts of co-authorship relations between countries based on the geographical location of each author's institution. Specifically, if a paper or book is affiliated with institutions from more than two countries, \emph{e.g.,} Norway, UK, and India, three co-authorship relations will be included in this dataset: Norway-UK, Norway-India, and UK-India (which could be regionally aggregated to one within Europe co-authorship relationship and two between Europe and Asia co-authorship relationships). Subsequently, we further aggregate the data into ten time periods: 1970-1974, 1975-1979, 1980-1984, 1985-1989, 1990-1994, 1995-1999, 2000-2004, 2005-2009, 2010-2014, and 2015-2018. The final time period does not include 2019 as the Scopus database for this year is as of yet incomplete\footnote{\added{However as we only apply methods within each time period, this missing year does not prevent us from considering this final period.}\deleted{The normalisation process used for network weights allows us to consider this period nonetheless, see Section 5.}}. 

\section{Trends in the global production of knowledge}

\begin{figure}[t!]
\centering
\includegraphics[width=.95\textwidth]{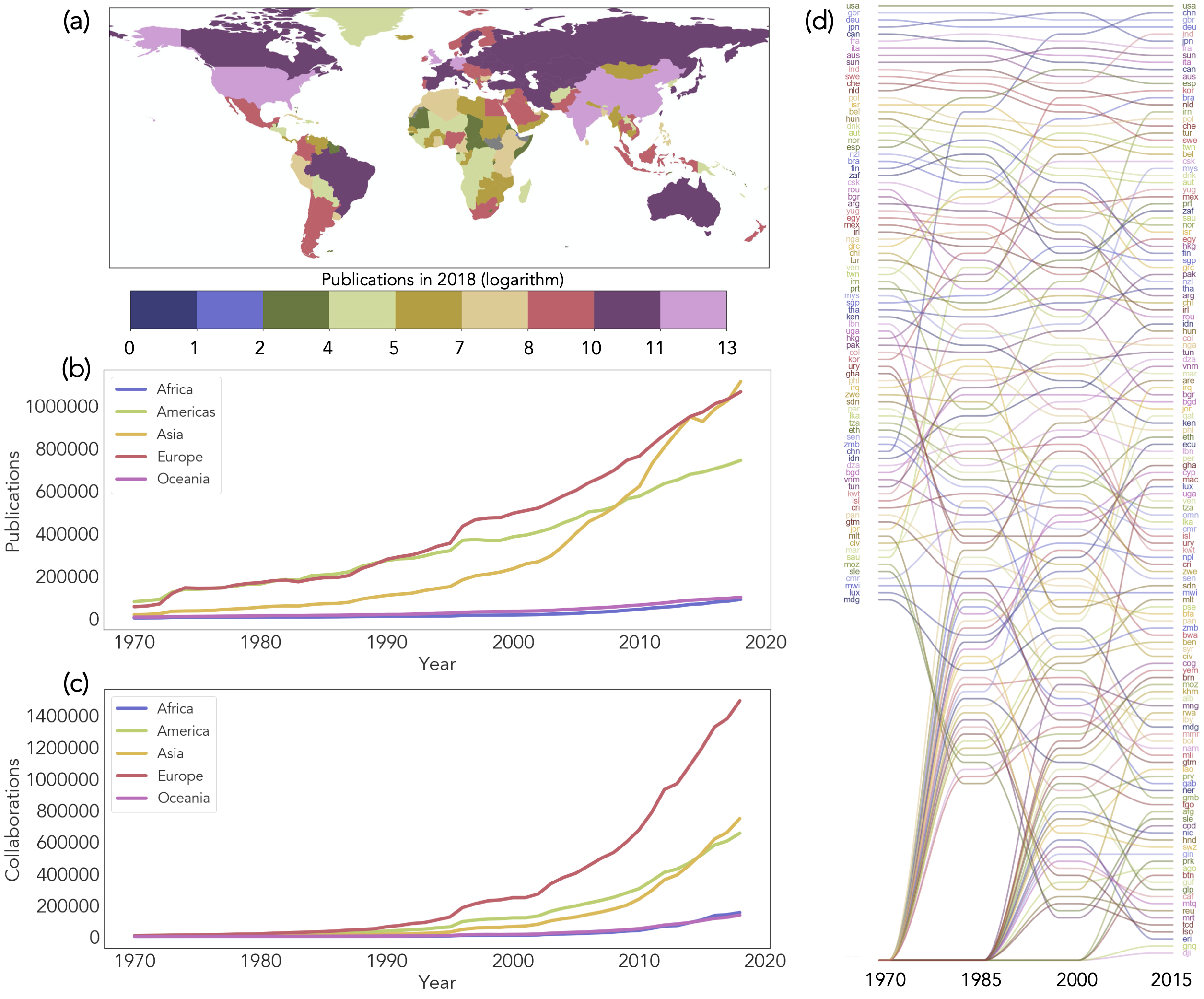}
\caption{Subfigure (a) maps the total number of academic publications in 2018 (log). Subfigures (b) and (c) show the temporal evolution of the total number of publications, and internationally coauthored publications, across geographical regions between 1970 and 2018. Finally, subfigure (d) displays the rank of countries in terms of the number of academic publications in 1970, 1985, 2000, and 2015. We observe some emerging economies such as China, Korea, Iran, and Malaysia rising significantly in rank towards the end of the time period.}
\label{fig:Descriptives}
\end{figure}

The production of academic publications is highly unequally distributed geographically. Figure~\ref{fig:Descriptives} (a) shows that the highest volume of publications is currently authored in the United States, United Kingdom, Germany, China, and India, while the lowest numbers can be found within Africa and Latin America. Looking back over the past five decades, Figure~\ref{fig:Descriptives} (b) reveals that Asia is catching up with Europe and the Americas, while the growth of academic publishing is much slower for Africa and Oceania. We contrast this with the growth of co-authored publications and find that growth was much faster in Europe than other continents, in particular after the turn of the century. While Asia is catching up to the Americas, international collaborations are growing much slower than its share of overall publications. 

Figure~\ref{fig:Descriptives}(d) displays the rank of countries in terms of the number of academic publications in 1970, 1985, 2000, and 2015. We observe that while countries in Europe and Asia as well as the United States and Canada are topping the list, some emerging economies such as China, Korea, Iran, and Malaysia significantly increased in rank towards the end of the time period\footnote{In the interest of readability, subfigure (d) omits any countries with less than 50 publications in one of the time periods, fewer than 100,000 inhabitants, and the group of small island developing nations (SIDS) except Singapore.}.

\begin{figure}[t!]
    \centering
    \includegraphics[width=0.9\textwidth]{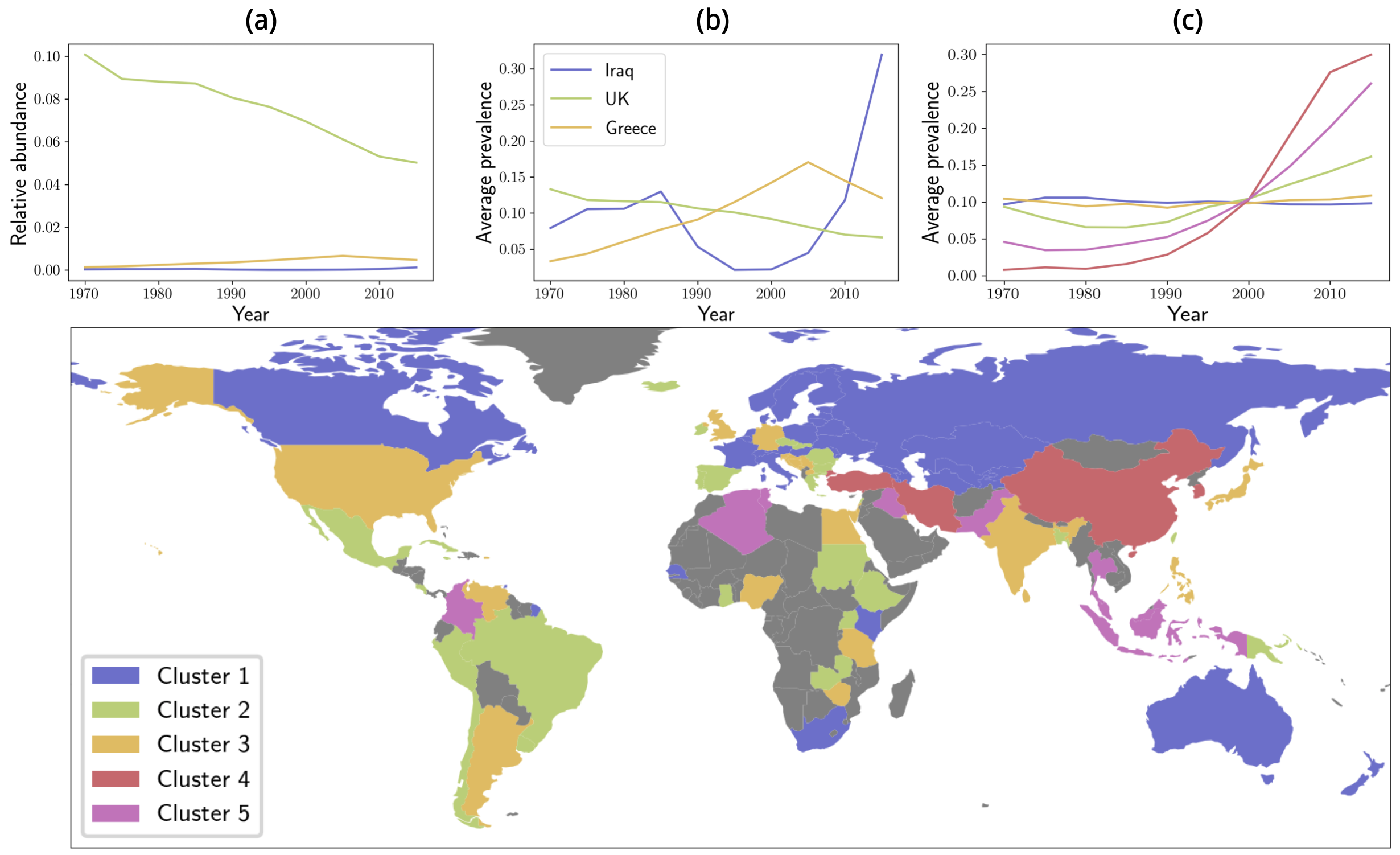}
    \caption{Subfigure (a) displays relative abundance profiles for Iraq, the UK and Greece. While the decline in relative share of global publications from the UK is clear, it is difficult to compare with other countries due to differing overall levels of production. Thus in subfigure (b) we display the average prevalence for the same three countries (in same colours as (a))---it is clear that while the relative share of the UK has slowly declined, for Greece it was increasing until 2010 (the subsequent decline may be due to the imposition of austerity). For Iraq we see a period of decline (during conflict) between 1990-2010, from which it is now recovering. In subfigure (c) we show the average prevalence of five groups of countries obtained from clustering their historical profiles. Two clusters display rapid growth in recent periods (red and purple), while two others display high variability (yellow and green). The final group (blue) features countries with stable or moderately declining profiles. Finally, in subfigure (d) we map these groups (in same colours as (c)), with the majority of the Global North belonging to stable or declining clusters while the Global South remains dynamic.}
    \label{fig:av_prev_full}
\end{figure}

It is clear that national publication and co-authorship rates have been subject to significant change over the past decades. Before proceeding to disentangle co-authorship patterns over time, we desire a simple method to systematically uncover which countries are emerging as research leaders in terms of publication growth (relative to size). To do so, we follow the method described in \citet{gargiulo_classical_2016}. First, we calculate the \emph{relative abundance} of publications of each country within each time-step. That is, at each time-step, we compute the global share of publication activity of country $i$:
\begin{equation}
    r_1(i,t)=\frac{n_i^{(t)}}{\sum_j n_j^{(t)}},
\end{equation}
where $n_i^{(t)}$ denotes the total number of publications produced by country $i$ in time period $t$. However, as shown in Figure~\ref{fig:av_prev_full} (a) for the countries Iraq, the UK, and Greece, it is a poor measure to compare the historical profiles of countries with dramatically different levels of production. To overcome this, we normalise each country's relative abundance profile by its total production across the full time period to obtain a measure of \emph{average prevalence}:
\begin{equation}
    r_2(i,t)=\frac{r_1(i,t)}{\sum_{t'} r_1(i,t')}.
\end{equation}
To ensure fair comparison, here we require each country to have produced more than 100 publications in each and every time period. Figure~\ref{fig:av_prev_full} (b) displays this metric for the same three countries, and now the relative trajectories of each country is clear: the UK has slowly declined in relative publication share, while Greece proportionally increased until 2010 (the subsequent decline may be due to the imposition of austerity). Iraq steeply declined in relative publication share from 1990 (possibly due to conflict after the invasion of Kuwait) and has only recovered more recently.

We then use these profiles to cluster countries with similar historical trajectories. We first calculate the Kolmogorov-Smirnov distance between each pair of country profiles (the supremum difference between the cumulative distribution of each profile \citep{smirnov_estimation_1939}), then use this distance matrix as the input for an agglomerative clustering algorithm. This algorithm works by first setting the maximum number of clusters to six (by looking at the corresponding clustering dendrogram), then finding the minimum threshold $r$ such that the distance between any two points within each cluster is less than $r$, and there are at most six clusters---see \emph{e.g.,} \cite{mullner_modern_2011}. These clustered profiles are displayed in Figure~\ref{fig:av_prev_full} (c), where each line corresponds to the average prevalence value of a cluster---note that as such Clusters 1 and 3 seem comparable, but the variance of profiles within Cluster 3 is much greater.

In Figure~\ref{fig:av_prev_full} (d) we display a map of the world coloured by these clusters. Five profiles are typical: the blue cluster (Cluster 1) corresponds to countries with reasonably stable profiles over the period investigated, such as Norway and much of the Global North. The green and yellow clusters (Clusters 2 and 3) include countries with periods of relative growth and decline, such as the UK and Greece. Finally, the red and purple clusters (Clusters 4 and 5) correspond to countries that have greatly increased their publication share in recent years such as Iraq. Amongst these, every region of the Global South has countries which have considerably improved their trajectory in recent times, from Colombia in South America to China in Asia.

The international research landscape is clearly undergoing continued structural change with new leaders emerging from all corners of the globe. Here we ask, how has this shift in the geographic spread and dynamics of knowledge production shaped a re-configuration of cross border collaboration ties?

\section{The dynamics of international vs regional collaboration diversity}

\vspace{0.3cm}
We wish to quantify how countries have changed their patterns of collaboration over time, focusing particularly on neighbouring and distant ties. One way to do this is to measure a countries' diversity of links to collaboration partners both within their own region and with countries in other regions.

In order to do this, we first calculate the Shannon entropy (see \emph{e.g.,} \cite{evans_community_2011, kumar_normalized_1986}) of the distribution of collaboration partners for each country. This provides us with a measure of the spread of collaborations for each country: values closer to one correspond to countries collaborating evenly with many countries around the world, and low values correspond to a narrow focus on collaboration with few countries. To be specific, we define the \emph{collaboration entropy} for country $i$ as
\begin{equation}
    CE(i,t)=-\frac{1}{\log (N-1)}\sum_{j\neq i} p_{ij}^{(t)}\log p_{ij}^{(t)} ,
\end{equation}
where $N$ is the total number of countries in our dataset in the time period\footnote{Note: again only countries which produced more than 100 total publications in all time periods are included here so as to ensure comparability across time.}, and
\begin{equation}
    p_{ij}^{(t)}=\frac{n_{ij}^{(t)}}{\sum_{j\neq i} n_{ij}^{(t)}},
\end{equation}
where $n_{ij}^{(t)}$ is the number of collaborations of academics from country $i$ with those in country $j$ in time period $t$. 

We are interested in investigating whether countries are collaborating more diversely within their region compared to outside their own region (continent). Hence, we decompose $CE$ as follows:
\begin{eqnarray}
    CE_{in}(i,t) &= &-\frac{1}{\log (N-1)}\sum_{j\in J_u} p_{ij}^{(t)}\log p_{ij}^{(t)} \\
    CE_{out}(i,t)&=&-\frac{1}{\log (N-1)}\sum_{j\in J_o} p_{ij}^{(t)}\log p_{ij}^{(t)} ,
\end{eqnarray}
where $J_u$ is the set of countries in the region of country $i$, and $J_o$ is the set of countries outside the region of country $i$. Diversity increasing within regions when compared to diversity between regions suggests stronger regional clustering, and impacts a variety of network measures analysed later. In particular, if the total strength of internal collaborations relative to external collaborations also increases (as verified in Appendix~\ref{app:strengths} and shown in Figure~\ref{fig:in_out_strengths}), it implies the formation of localised regional collaboration networks, or knowledge basins within which knowledge circulates more easily. 

\begin{figure}[t!]
    \centering
    \includegraphics[width=0.9\textwidth]{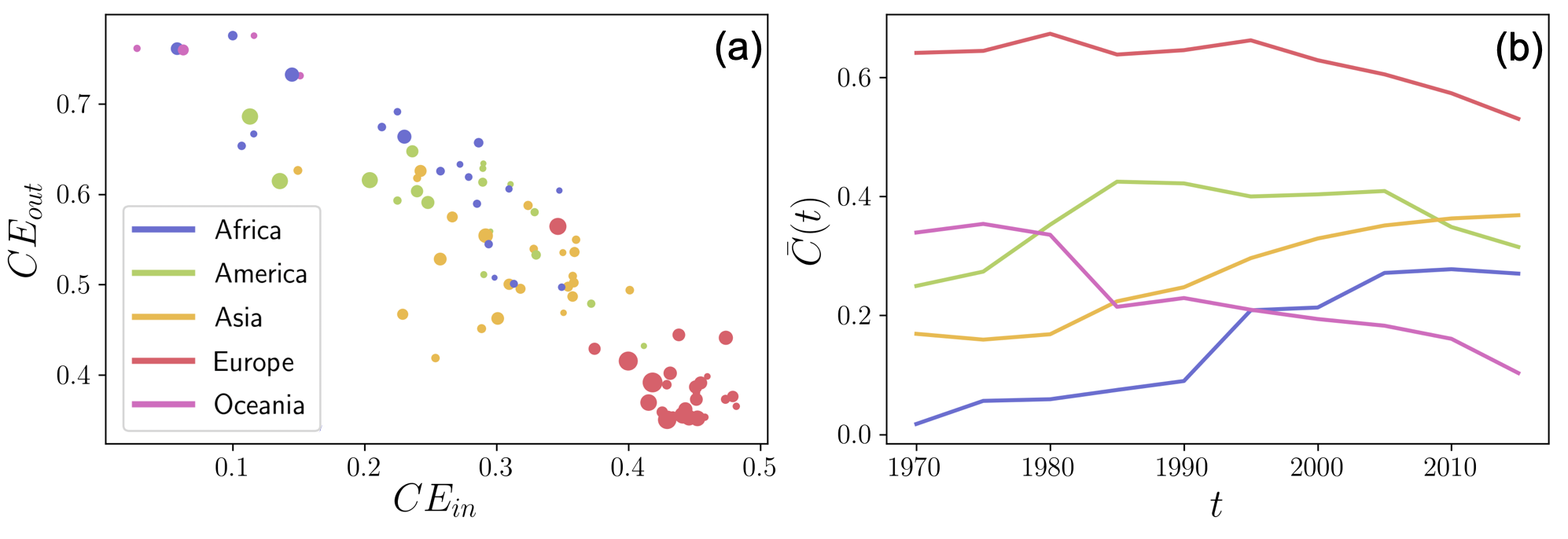}
    \caption{Subfigure\deleted{s} (a) \deleted{and (b)} display\added{s} the collaboration \deleted{and partner entropy metrics}\added{entropy measure} within and outside of the region of each country in the final time period, with points scaled by the total number of collaborations. We observe that \added{European countries tend to have diverse collaborations within their region relative to those further afield, while for many countries in Africa or the Americas the reverse is true}\deleted{some countries such as Italy and Senegal strongly favour diverse partnerships within their region, while others like France and the UK prioritise those outside their region}. Subfigure\deleted{s} (\deleted{c}\added{b}) \deleted{and (d) }plot\added{s} the average proportion of within-region entropy (as a share of total entropy) for each continent. We observe that Africa and Asia have greatly increased their focus on diverse within-region collaboration since 1970.}
    \label{fig:full_ent}
\end{figure}

We plot $CE_{in}$ versus $CE_{out}$ for the final time period for all countries in Figure~\ref{fig:full_ent} (a), where the size of the points is scaled by the total number of collaborations. We observe that European countries (shown in red) seem to collaborate more diversely with each other than with the rest of the world, while for many African countries (shown in blue) the reverse seems to be the case. 

\deleted{However, it may not be surprising that Europe collaborates more internally due to the higher overall regional capacity for knowledge production, and vice versa for Africa. }
\deleted{Hence, we compute a second form of entropy intended to measure the diversity of significant collaboration partners while accounting for publication capacity.}

In order to assess the dynamics of inter- and intra-region collaboration diversity over time, we compute the proportion of within-region entropy (as a share of total entropy) for each country: 
\begin{eqnarray}
    C(i,t) & =&\frac{CE_{in}(i,t)}{CE_{in}(i,t)+CE_{out}(i,t)}. 
\end{eqnarray}
We plot the mean value---across countries in a region---of \added{this quantity}\deleted{these quantities} over time in Figure~\ref{fig:full_ent}\added{(b)}\deleted{(c) and (d)}. 
\deleted{Focusing on strong partnerships as captured by partnership entropy on the right, we}\added{We may} observe that within-region diversity was high but has been \deleted{steadily}\added{slowly} declining in Europe \added{since 1995, suggesting the region is broadening its focus to some extent}. On the other hand, within-region collaboration diversity increased significantly for the Americas and Asia from the 1980s, and for Africa after 1990. However, there appears to be a general small decline in within-region diversity (relative to out-of-region collaboration diversity) in the final two time periods \added{for the Americas}, suggesting a recent opening up of \added{their}\deleted{global} collaboration networks. \deleted{An analysis of the sensitivity of these results to random perturbations of the original (count) data is conducted in Appendix~\ref{app:ent-sens}, and demonstrates the robustness of these observations.}

\section{The evolving structure of research clusters in the global collaboration network}

The change in research focus, from international to regional collaboration, observed in the previous section provokes a more general investigation of how knowledge flows (as proxied by academic collaborations) may have changed over time. In particular, we ask whether these trends have translated into an overall consolidation of regional ties, creating isolated clusters or pools of knowledge production.  

To uncover the complex structure of these flows, we construct a network where the nodes are countries, and the edges correspond to \added{the number of collaborations between countries $i$ and $j$ at time $t$, $n_{ij}^{(t)}$, such that the network at this time has adjacency matrix, $A^{(t)}$, with the corresponding $i$, $j$th entries.} \deleted{collaboration strength (as defined equation~\eqref{eqn:network}). This network has adjacency matrix $A=\hat{P}^{(t)}$ at time $t$, with entries given by equation~\eqref{eqn:network}.}

\added{Prior to further analysis, to immediately visualise significant partnerships, we follow a similar procedure to that proposed by} \citet{neffke_skill_2013} \added{for estimating skill-overlap between industry pairs based on inter-industry job transitions. The logic behind doing so is similar to that for revealed comparative advantage (RCA, see \emph{e.g.} }\cite{balassa1965trade}\added{), in that measures calculated on the network formed by raw counts are typically dominated by those locations with the highest overall production (\emph{i.e.} USA, China and similar). Instead, we normalise the observed counts by the capacity of each country, measured by total collaborations, using a configuration model-like approach, apply a transformation to help account for the spread of subsequent results, and finally apply a thresholding step. The details may be found in Appendix~\ref{app:weight_trans}.}

In Figure~\ref{fig:Networks_heatmaps}, we display this \added{transformed} network, \added{with edges with strengths given by equation~\eqref{eqn:network},} for the five-year periods commencing in 1970, 1985, 2000, and for 2015-2018, where countries which belong to the same continent have the same colour, and the size of each country is proportional to their total number of publications within that time period. The spring algorithm ForceAtlas in Gephi is used to layout each network, and edges above a 0.5 threshold are shown. 

We observe that countries tend to cluster together geographically in latter time periods. This can be seen with respect to the United Kingdom and Germany: in earlier time periods they occupy fairly central `positions', but in the latter time periods locate more closely to other European countries. On the other hand, while we note that the rise of publication volume in China and India is visible particularly over the past two decades, the positions of these countries along with Japan remain relatively close to their regional groups. 
In Figure~\ref{fig:Networks_heatmaps}(e), we display the mean edge weights between the five continents in the form of aggregated adjacency matrices. We observe the emergence of a defined diagonal from the year 2000, while the off-diagonals grow paler. This indicates that intra-regional collaborations have strengthened, while the inter-continental collaborations appear to decline. Once again, we observe that in the most recent time period this trend may be beginning to change, with more intercontinental partnerships emerging. 

\begin{figure}[t!]
\centering
\includegraphics[width=.95\textwidth]{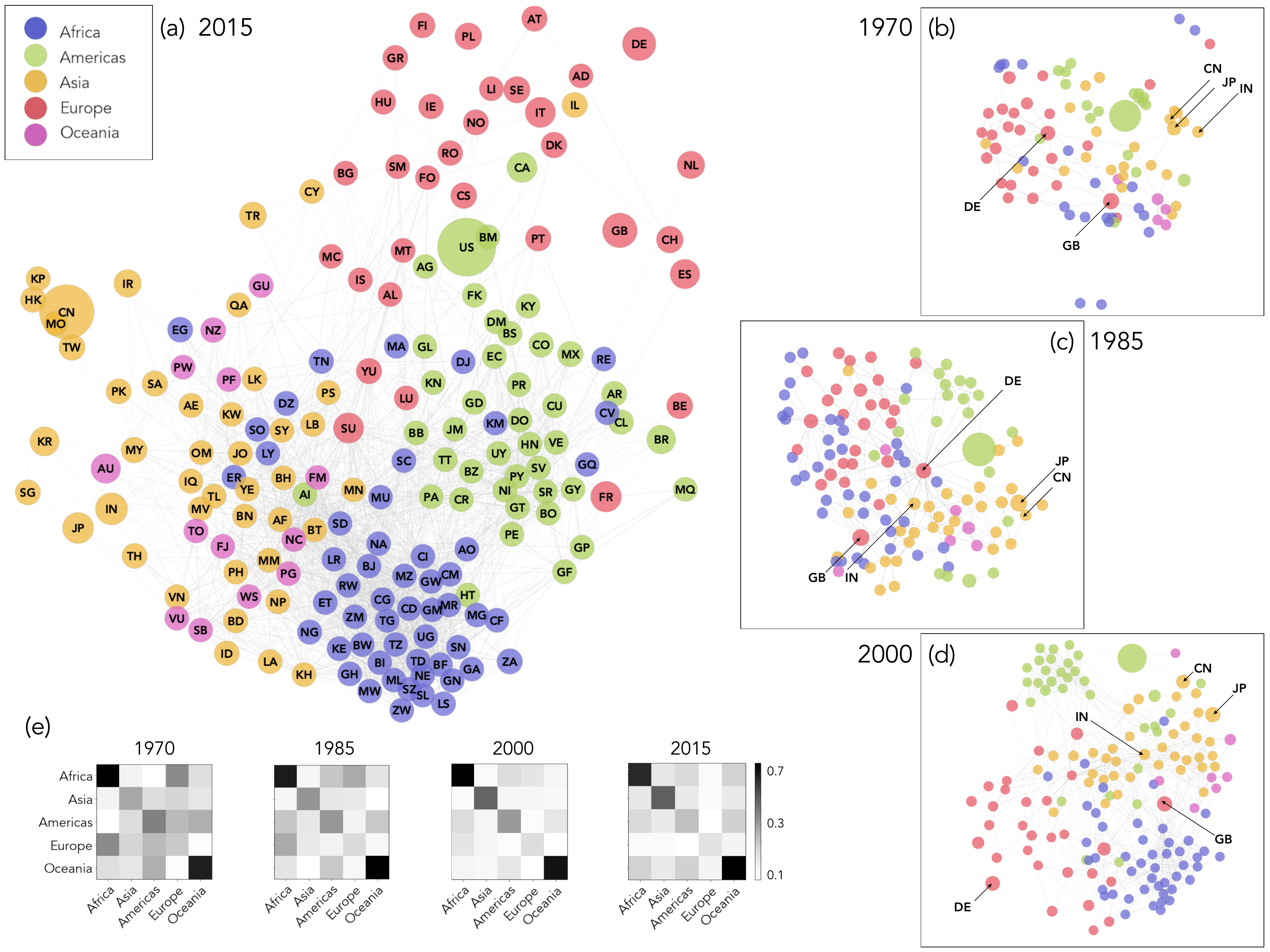}
\caption{Subfigure (a) shows the global collaboration network constructed for data between 2015-18. The algorithm ForceAtlas is used to layout the network, with edges above a 0.5 edge-weight threshold shown. Node sizes are scaled according to the total number of publications during this period, labeled with the ISO2 code corresponding to each country, and coloured according to geographical regions. Subfigures (b)-(d) show analogous networks constructed for periods 1970-74, 1985-89, and 2000-04. Finally, subfigure (e) displays mean edge weights between the five continents in the form of adjacency matrices. We observe a concentration of edge weights within regions, as seen by the shaded diagonal entries, particularly from 2000. Colours correspond to the same continent in each plot.}
\label{fig:Networks_heatmaps}
\end{figure}

In order to explore the increasing `regionalisation' of research collaboration, we wish to extract information from the networks about groups of countries engaged in intense research collaboration across time. Exploring such groupings is a key focus of network science, known as \emph{community detection}. Loosely speaking, this corresponds to a partition of nodes into communities for which within-community links are significantly stronger that between-community links. It is often found to be the case that these naturally arise in the real world, \emph{e.g.} in social, neurological, or indeed academic networks as under consideration here \citep{newman_finding_2004}. Here, such communities reveal groupings of countries which engage in significant research collaboration -- and analysis of their evolution over time enables us to extract a quantitative description of the changing global research landscape.

While a variety of methods exist (see \emph{e.g.} \cite{javed_community_2018} for an overview), the approach we take is that of \added{optimisation of linearised stability} \citep{delvenne2010stability, lambiotte2008laplacian, lambiotte2011flow}\deleted{optimisation of Newman-Girvan modularity }
. 
Given a partition $X$, this \deleted{metric}\added{method involves} comput\added{ing}\deleted{es} \added{a} sum of the deviations of the network \added{edges within each} communit\added{y}\deleted{ies} from a \added{weighted} \emph{configuration null model} (where edges are shuffled randomly but node strengths are preserved). Mathematically, 
\begin{equation} \label{confmod1}
 	Q_{\text{conf}}(X) = \frac{1}{2m}\sum_{i, j} \left\{A_{ij} - \gamma\frac{k_{i}k_{j}}{2m} \right\}\delta(x_{i}, x_{j}) ,
\end{equation}
where 
\begin{equation}
 	k_{i} = \sum_{j} A_{ij}, \quad 2m = \sum_{i} k_{i} ,
\end{equation}
are respectively the strength of node $i$ and total edge weight of the network, $x_{i}$ is the community of node $i$ (thus $\delta(x_{i}, x_{j}) = 1$ if $i$ and $j$ are in same community and is zero else), \added{and $\gamma$ is a so called `resolution' parameter. This final parameter controls the contribution of the null model to the sum, and so affects which partition will be optimal -- larger values favour recovering smaller communities, and \emph{vice versa}. Under the configuration null model, the expected strength of link between $i$ and $j$ is $k_{i}k_{j}/2m$---\emph{i.e.,} the total strength of node $j$ times the probability of connecting to node $i$. In particular, using this null model, if $\gamma=1$ linearised stability is identical to the conventional Newman-Girvan modularity} \citep{newman_finding_2004}. \added{This linearised form is also effectively identical to another method previously introduced in} \cite{reichardt2006statistical} \added{for modularity at different network scales.} \added{This tuning parameter is highly useful, as it allows us to avoid to some extent the resolution limit that typical modularity has been shown to face} \citep{fortunato_resolution_2007}\added{, in that it is possible to fail to detect non-trivial small communities.} 

\added{The principal idea behind stability is that if we follow walkers around the network, which jump between nodes with a probability proportional to the edge weight, then over time sets of nodes where walkers spend a prolonged period suggest denser connections within such a set than to outside, \emph{i.e.} they form form a community. The period of time for which we track such walkers naturally leads to the resolution parameter $\gamma$. More details on this are provided in Appendix~\ref{app:stability}.}



In order to find a node partition $X$ which maximises this function, a typical approach is to use a greedy algorithm by \citet{blondel_fast_2008}. This works by initially placing each node in its own community, then iteratively merging nodes with those adjacent to themselves if an increase in \added{linearised stability}\deleted{modularity} is achieved. This process is stochastic in the sense that it may produce a slightly different optimum partition depending on the order in which nodes are `visited'. It is efficient as only local information (nearest neighbours) to the node is necessary at each step. Recently there has been a further improvement with a similar logic, known as the Leiden algorithm \citep{traag_louvain_2019}: this appears to result in higher \added{linearised stability}\deleted{modularity} with lower computational cost, and so will be used here. \added{Through studying the variation of information (a metric for comparing partitions) as described in Appendix~\ref{app:stability}, we find that two resolution times $\tau=1/\gamma$ of interest are $\tau=1.0$ (\emph{i.e.} actually conventional modularity) and $\tau=0.76$, which provides a finer-grained view of the network.}

\begin{figure}[t!]
    \centering
    \includegraphics[width=0.9\textwidth]{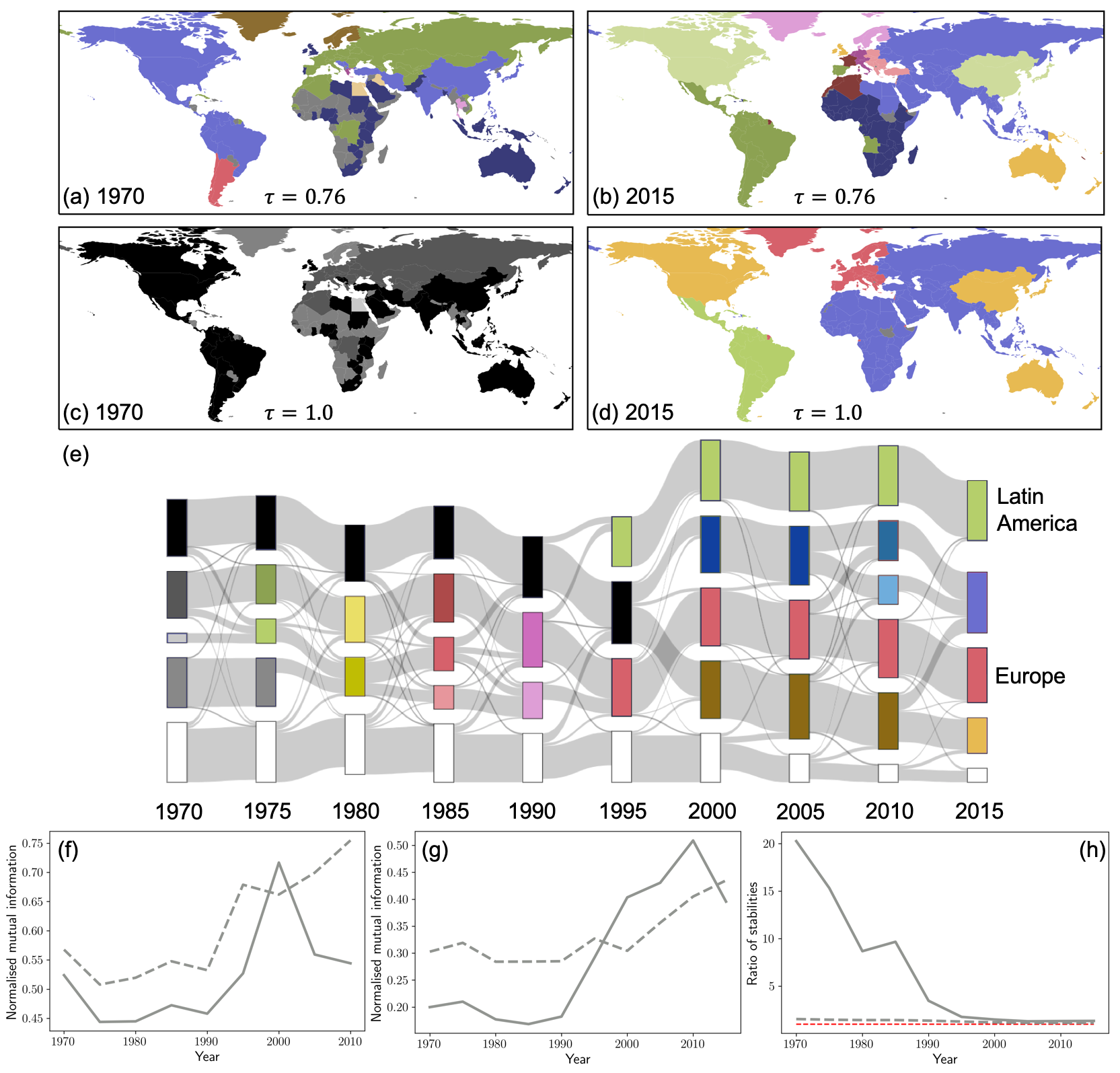}
    \caption{\added{In subfigures (a) and (b), we display the communities found in the global collaboration network constructed from data between 1970-74 and 2015-18 respectively,  using $\tau=1/\gamma=0.76$. We can observe for instance that in 2015, Europe in particular appears highly (sub-)regional, and much of Oceania becomes its own separate community at this finer scale. Subfigures (c) and (d) show communities in these same periods, coloured as in the Sankey plot in (f) below, but found instead with $\tau=1.0$}. We observe that while in 1970 \added{these} communities \deleted{of significant partnerships} were globally distributed (in some cases along colonial lines), in 2015 they appear to \added{overall} be more regionally focused. This is supported by subfigure (e), where communities are connected between adjacent periods if their Jaccard (similarity) index is greater than 0.6. Subfigure (f) compares the community structure of adjacent time periods using the NMI, showing \added{generally} increasing stability of the community structure over time. Subfigure (g) compares each partition to a partition where nodes are split into continents, highlighting the increasing similarity of the communities to continents over time. Finally, subfigure (h) displays the ratio of the \added{stability}\deleted{modularity} of each partition to the continental partition, \added{with the dashed red line at 1 thus corresponding to equal stability}, further supporting the latter insight. \added{For each of subfigures (f), (g) and (h), results for communities found with $\tau=1.0$ are shown solid, and dashed for $\tau=0.76$.}}    
    \label{fig:full_comms_fig}
\end{figure}

We display the \added{best} partitions $X^{(t)}$ found from applying this optimisation process\added{, with $\tau=1.0$,} to the network constructed for each time period in Figure~\ref{fig:full_comms_fig}(e). Following a similar approach to that of \citet{pietilanen_dissemination_2012} and \citet{fagan_assessing_2018}, 'flows' between two communities $A$ and $B$ are scaled according to the Jaccard index $J(A,B)=|A\cap B|/|A\cup B|$. We first assign each community a colour arbitrarily, then compare adjacent time periods and retain the previous colour if $J(A,B)>0.6$. The white community corresponds to countries outside of the time period under consideration. 
This figure contains a wealth of information, for instance evidencing that \deleted{significant partners}\added{collaboration patterns often} changed more regularly in earlier, more turbulent decades, before beginning to settle from 1995 onwards. \added{It may be seen for example that Europe consolidates as a block at this scale from 1995 onwards, shortly after the formation of the European Economic Area (EEA).} We observe that in the final time period, there are \deleted{five}\added{four} communities which roughly correspond to the regions of \deleted{Africa, Europe, Asia, the Americas and the Middle East}\added{Europe and Latin America, North America with China, Australia and nearby countries, and the rest of the world. The community of North America \emph{et al.} may be an artefact of the USA and China being the two major global producers, and suggests that an alternative null model could be more suitable depending on the goal of analysis -- we explore the deviation from the null model further in Appendix~\ref{app:kl-div}, but leave the development of such an alternative to future work.}

In order to further investigate the rate of change of the modular structure over time, and the observed `regionalisation' of research collaboration ties, we wish to quantify the similarity between each partition and its preceding partition, and between each partition and the `continental partition' (where countries are assigned to a community based on continent).
While the Jaccard index is good measure for comparing pairs of communities, to compare partitions we instead calculate the normalised mutual information (also known as the symmetric uncertainty \citep{witten_data_2002}). This is defined by
\begin{equation}
   NMI(X,Y)=2\frac{\sum_{i,j} r_{ij}\log\left(r_{ij}/p_i q_j\right)}{\sum_k p_k\log p_k + \sum_\ell q_\ell \log q_\ell},
\end{equation}
for two partitions $X$ and $Y$, where $n$ is the number of nodes, and $p_{i}=|X_{i}|/n$ (the share of nodes in community $i$ of $X$), $q_{i}=|Y_{i}|/n$ (the share of nodes in community $i$ of $Y$), and $r_{ij}=|X_{i}\cap Y_{j}|/n$ (the share of nodes in both community $i$ of $X$, and community $j$ of $Y$). 

We compare the partitions obtained in adjacent time-steps through calculating the normalised mutual information: \emph{i.e.} $NMI(X^{(t)},X^{(t+1)})$, where $X^{(t)}$ is the partition obtained for time period $t$. \added{In} Figure~\ref{fig:full_comms_fig} (f), \added{we display the values of this function over time at two different scales, with $\tau=1.0$ shown solid, and the finer scale $\tau=0.76$ shown dashed. We observe} \deleted{shows} that after an initial period of change, recent years have seen relatively stable global research communities form \added{at the finer scale, while at the more aggregate scale there is still some change (primarily due to splits in the large, `rest of the world', community shown in purple)}. Next, we construct a new partition, $C$, which divides the world into five continents (communities): each country is assigned to their continent, \emph{i.e.} Africa, America, Asia, Europe, and Oceania. In order to see how similar each partition is to this continental partition, we calculate $NMI(X^{(t)},C)$ for all $t$. Figure~\ref{fig:full_comms_fig} (g) confirms what we had suspected from previous figures in that there has been a clear trend towards regionalisation of research ties \added{at both scales}, particularly between 1990-2010.  

As a final check, we compare the \added{stability}\deleted{modularity} of each detected partition to the \added{stability}\deleted{modularity} of the continental partition. Since \added{stability}\deleted{modularity} is a measure of partition quality, we would expect the \added{stability}\deleted{modularity} of the continental to approach that of the detected partition in latter time periods. It is important to understand the difference in quality between these partitions, particularly as there is inherent randomness to the optimisation algorithm used, and it only guarantees convergence to a local optima. In other words, the `optimal' partition we find could in fact be only marginally better than the continental partition in early decades, even if the partitions themselves were very different as measured by NMI. We cannot compare raw values of \added{stability}\deleted{modularity} across time, as it varies with respect to network size/density etc.---as such, we compute the ratio
\begin{equation}
    Q_{rat}(X^{(t)},C)=\frac{Q_{conf}(X^{(t)})}{Q_{conf}(C)}.
\end{equation}
The ratio of the \added{stability}\deleted{modularity} scores tells us how well the geographic (or continental) partition `performs' as a set of communities compared to those detected by our community detection algorithm. 
Figure~\ref{fig:full_comms_fig} (h) confirms that, as expected, this ratio declines over time. More specifically, we observe that the continental partition was of significantly lower quality in earlier time periods, \added{particularly for the scale with $\tau=1.0$,} suggesting this was not a good `description' of the network structure at that time. 
In later periods, the ratio approaches 1 \added{(shown dashed red) at both scales}, suggesting that the continental partition is increasingly a good fit for the network structure.

\section{Discussion}

The creation and diffusion of knowledge between nations is crucial for the advancement of skills and capabilities, critical drivers of economic development. Patterns of knowledge diffusion via research collaboration on a global level, however, remain poorly understood. We address this research gap through analysing a worldwide dataset of international scientific publications spanning all major disciplines over five decades. We find that collaboration ties appear to have become more localised since 2000, with researchers prioritising regional co-authorship relative to more distant ties. We corroborate this insight via an analysis of the evolving modular structure of the global collaboration network, finding a recent stabilisation of research clusters along \added{increasingly} regional lines.

These findings were unexpected given the generally accepted wisdom on the onward march of globalisation, and thus have a number of significant implications. 
On one hand, this could be a positive signal: research expertise is growing in many previously under-equipped nations and regions, and hence scholars no longer have to look as far afield as they once did. 
Regional research efforts may be driven by resident researchers focusing their efforts  on addressing particular economic, social, and political concerns within the region. Specific research programmes have been introduced to strengthen scientific collaboration within regions such as the Horizon 2020 (soon to give way to Horizon Europe) initiative in Europe. 
Further, regional research and development programmes increasingly make use of the `smart specialisation' model in research, whereby countries with well-defined domains of specialisation (e.g., in research and innovation) are seen as more likely to produce research excellency in specific areas. 
These countries are then chosen as sites for related regional research programmes and institutes, with the aim of anchoring and nurturing these localised sites of expertise. This model was originally developed by the European Union in order to address a transatlantic gap in R\&D but has since been adopted by many regions and countries \citep{gomez_prieto_smart_2019}---a trend that our research findings would seem to support and perhaps a driving force behind some of the patterns 
we have identified. 
On the other hand, such a retrenchment may be worrying, given what we know about the importance of capability building and knowledge diffusion through `on-the-job' learned experience, leading to an uneven distribution of capabilities across regions. 
Indeed, the role of donors in strengthening local research capacity through international collaborations in many lower and middle-income economies has been deemed crucial, as these countries tend to lack research capacity and face problems translating research into impact. 
In this respect, it seems more important than ever that large research funders such as those within the EU and the US support international collaboration on a scale that far outstrips current levels. 
While funders, such as the US Agency for International Development's (USAID) Partnerships for Enhanced Engagement in Research or the UK's Newton Fund already have dedicated mechanisms to support North-South research partnerships, this could mean expansion of National Science Foundation (NSF) programmes to allow non-US research leads, or a U-turn in the recent decrease in funding allocated to the much-feted Fulbright programme (which supported two of the authors of this paper to spend time in the US). 
One bright spot is the recent growth of development-oriented research funding in the UK, the Global Challenges Research Fund (which supported this work), that not only supports equitable UK-developing nation collaboration but mandates it. It is only with large scale investment in such programmes that international research collaboration will continue to play a vital role in global capability building.

While previous work comparing data sources on academic publishing highlight the comparative strength of the Scopus dataset, we are aware that there are limitations to this data given our interest in comparison between countries. The database's coverage is thought to be weaker for the social sciences and humanities, and for literatures in other languages than English \citep{aksnes_criteria-based_2019}. Furthermore, we cannot ascertain that the indexing of work from publishers located in countries where academia is less well resourced is as complete as for countries with more established academes. Additionally, our dataset includes journal articles, books, and conference proceedings, but no other types of academic outputs. Ideally one would complement this analysis with additional material from academic search engines such as Google Scholar, which contains up to four times as many documents as Scopus. However, due to well-known issues such as document duplication, false citations, and unstable search indexing, this would require a major investment in data cleaning and processing. 

There are a few clear avenues for future work. Firstly, much remains to be understood about the nature and evolution of global collaboration networks, and the roles of individual actors. For example, it would be interesting to investigate whether we can identify global hegemons, countries playing the role of gatekeeper between lesser regional partners and the rest of the world. Similarly, our work suggests that colonial links and geopolitical alliances shaped, for a time, regional basins of knowledge. Has this transition from historical blocs to regional clusters positively (or negatively) impacted the research capacity of less developed nations? In other words, who have been the winners and losers from this shift in network structure? 

Finally, there is ample scope and reason to further investigate the structure of global research ties and knowledge diffusion beyond inter-country links. First and foremost, research quality and disciplinary focus is often highly institution---rather than country---specific. Furthermore, funding programmes often target specific fields and institutions. For these reasons amongst others, it would be fruitful to dis-aggregate the global collaboration network by institution and field. Perhaps collaborations in certain disciplines are heavily demarcated along regional lines, while others flourish under international collaboration. Perhaps top-tier institutions maintain international links, while second-tier institutions focus more on regional ties. Additionally, there are a large number of possible metrics one might compare to co-authorship ties, including researcher mobility patterns. I.e, is the recent regional retrenchment in collaboration patterns also observed in researcher mobility patterns?


\begin{backmatter}

\section*{Abbreviations}

ASE: Academic Search Engine
IHERD: Organization for Economic Co-operation and Development Project on Innovation, Higher Education and Research \& for Development
NSF: National Science Foundation
OECD: Organization for Economic Co-operation and Development
RD: Research and development
UNESCO: United Nations Educational, Scientific and Cultural Organization

\section*{Declaration}

\subsection*{Acknowledgements}

This work uses Scopus data provided by Elsevier through the International Center for the Study of Research (ICSR) Lab.

\subsection*{Funding}

This publication is based on work partially supported by the EPSRC Centre For Doctoral Training in Industrially Focused Mathematical Modelling (EP/L015803/1) in collaboration with Elsevier (John Fitzgerald).

Sanna Ojanper\"a is currently funded by The Alan Turing Institute doctoral grant TU/C/000020.

This article was completed with support from the PEAK Urban programme, funded by UKRI's Global Challenge Research Fund, Grant Ref: ES/P011055/1 (Neave O'Clery).

\subsection*{Availability of data and materials}

The data that support the findings of this study are available from Elsevier (Scopus) but restrictions apply to the availability of these data, which were used under license for the current study, and so are not publicly available. Data are however available from the authors upon reasonable request and with permission of Elsevier.

\subsection*{Author's contributions}

NOC, JF and SO designed the study and the methodology, JF and SO processed and analysed the results.

\subsection*{Ethics approval and consent to participate}

Not Applicable.

\subsection*{Consent for publication}

Not Applicable.

\subsection*{Competing interests}

The authors declare that they have no competing interests.

\bibliographystyle{unsrtnat}

\bibliography{references.bib,refs.bib,Sanna_refs.bib,final_refs.bib} 

\begin{thebibliography}{92}
\providecommand{\natexlab}[1]{#1}
\providecommand{\url}[1]{\texttt{#1}}
\expandafter\ifx\csname urlstyle\endcsname\relax
  \providecommand{\doi}[1]{doi: #1}\else
  \providecommand{\doi}{doi: \begingroup \urlstyle{rm}\Url}\fi

\bibitem[Powell and Snellman(2004)]{powell_knowledge_2004}
Walter~W. Powell and Kaisa Snellman.
\newblock The {Knowledge} {Economy}.
\newblock \emph{Annu. Rev. Sociol.}, 30\penalty0 (1):\penalty0 199--220, 2004.
\newblock ISSN 0360-0572.
\newblock \doi{10.1146/annurev.soc.29.010202.100037}.

\bibitem[Jackson(2011)]{jackson_what_2011}
Deborah Jackson.
\newblock What is an innovation ecosystem.
\newblock Technical Report 1(2), National Science Foundation, 2011.

\bibitem[{UNESCO}(2015)]{united_nations_educational_scientific_and_cultural_organization_unesco_2015}
{UNESCO}.
\newblock {UNESCO} {Science} {Report}: towards 2030.
\newblock Technical report, Imprimerie Centrale, Luxembourg, 2015.
\newblock URL \url{https://unesdoc.unesco.org/ark:/48223/pf0000235406}.

\bibitem[Hidalgo et~al.(2007)Hidalgo, Klinger, Barab{\'a}si, and
  Hausmann]{hidalgo_product_2007}
C.~A. Hidalgo, B.~Klinger, A.-L. Barab{\'a}si, and R.~Hausmann.
\newblock The {Product} {Space} {Conditions} the {Development} of {Nations}.
\newblock \emph{Science}, 317\penalty0 (5837):\penalty0 482--487, July 2007.
\newblock ISSN 0036-8075, 1095-9203.
\newblock \doi{10.1126/science.1144581}.
\newblock URL \url{https://science.sciencemag.org/content/317/5837/482}.
\newblock Publisher: American Association for the Advancement of Science
  Section: Research Article.

\bibitem[Frenken and Boschma(2007)]{frenken2007theoretical}
Koen Frenken and Ron~A Boschma.
\newblock A theoretical framework for evolutionary economic geography:
  Industrial dynamics and urban growth as a branching process.
\newblock \emph{{Journal of Economic Geography}}, 7:\penalty0 635--649, 2007.

\bibitem[{Hidalgo} and {Hausmann}(2009)]{HidalgoHausmann2009}
C{\'e}sar~A {Hidalgo} and Ricardo {Hausmann}.
\newblock The building blocks of economic complexity.
\newblock \emph{Proceedings of the National Academy of Sciences}, 106\penalty0
  (26):\penalty0 10570--10575, 2009.

\bibitem[{Hausmann} and {Hidalgo}(2011)]{hausmann2011network}
Ricardo {Hausmann} and C{\'e}sar~A {Hidalgo}.
\newblock The network structure of economic output.
\newblock \emph{Journal of Economic Growth}, 16\penalty0 (4):\penalty0
  309--342, 2011.

\bibitem[O'Clery et~al.(2019{\natexlab{a}})O'Clery, Curiel, and
  Lora]{oclery_commuting_2019}
Neave O'Clery, Rafael~Prieto Curiel, and Eduardo Lora.
\newblock Commuting times and the mobilisation of skills in emergent cities.
\newblock \emph{Applied Network Science}, 4\penalty0 (1):\penalty0 118,
  2019{\natexlab{a}}.

\bibitem[Ojanper{\"a} et~al.(2017)Ojanper{\"a}, Graham, Straumann, Sabbata, and
  Zook]{ojanpera_engagement_2017}
Sanna Ojanper{\"a}, Mark Graham, Ralph Straumann, Stefano~De Sabbata, and
  Matthew Zook.
\newblock Engagement in the {Knowledge} {Economy}: {Regional} {Patterns} of
  {Content} {Creation} with a {Focus} on {Sub}-{Saharan} {Africa}.
\newblock \emph{Information Technologies \& International Development},
  13\penalty0 (0):\penalty0 33--51, March 2017.
\newblock ISSN 1544-7529.
\newblock URL \url{http://itidjournal.org/index.php/itid/article/view/1479}.

\bibitem[Ojanper{\"a} et~al.(2019)Ojanper{\"a}, Graham, and
  Zook]{ojanpera_digital_2019}
Sanna Ojanper{\"a}, Mark Graham, and Matthew Zook.
\newblock The {Digital} {Knowledge} {Economy} {Index}: {Mapping} {Content}
  {Production}: {The} {Journal} of {Development} {Studies}: {Vol} 55, {No} 12.
\newblock \emph{The Journal of Development Studies}, 55\penalty0 (12):\penalty0
  2626--2643, 2019.
\newblock \doi{10.1080/00220388.2018.1554208}.
\newblock URL
  \url{https://www.tandfonline.com/doi/full/10.1080/00220388.2018.1554208}.

\bibitem[{Programme on Innovation, Higher Education and Research} and {for
  Development
  (IHERD)}(2012)]{programme_on_innovation_higher_education_and_research_research_2012}
{Programme on Innovation, Higher Education and Research} and {for Development
  (IHERD)}.
\newblock Research {Universities}: {Networking} the {Knowledge} {Economy},
  2012.

\bibitem[Wagner-D{\"o}bler(2001)]{wagner-dobler_continuity_2001}
Roland Wagner-D{\"o}bler.
\newblock Continuity and {Discontinuity} of {Collaboration} {Behaviour} since
  1800 --- from a {Bibliometric} {Point} of {View}.
\newblock \emph{Scientometrics}, 52\penalty0 (3):\penalty0 503--517, 2001.
\newblock ISSN 0138-9130.
\newblock \doi{10.1023/A:1014208219788}.

\bibitem[Meyer and Bhattacharya(2004)]{meyer_commonalities_2004}
Martin Meyer and Sujit Bhattacharya.
\newblock Commonalities and differences between scholarly and technical
  collaboration.
\newblock \emph{Scientometrics}, 61\penalty0 (3):\penalty0 443--456, 2004.
\newblock ISSN 0138-9130.
\newblock \doi{10.1023/B:SCIE.0000045120.04489.80}.

\bibitem[Narin et~al.(1991)Narin, Stevens, and Whitlow]{narin_scientific_1991}
F.~Narin, K.~Stevens, and Edith Whitlow.
\newblock Scientific co-operation in {Europe} and the citation of
  multinationally authored papers.
\newblock \emph{Scientometrics}, 21\penalty0 (3):\penalty0 313--323, 1991.
\newblock ISSN 0138-9130.
\newblock \doi{10.1007/BF02093973}.

\bibitem[Wagner and Leydesdorff(2005{\natexlab{a}})]{wagner_mapping_2005}
Caroline~S. Wagner and Loet Leydesdorff.
\newblock Mapping the network of global science: comparing international
  co-authorships from 1990 to 2000.
\newblock \emph{International Journal Of Technology And Globalisation},
  1\penalty0 (2):\penalty0 pp185--208, 2005{\natexlab{a}}.
\newblock ISSN 1476-5667.

\bibitem[Wuchty et~al.(2007)Wuchty, Jones, and Uzzi]{wuchty_increasing_2007}
Stefan Wuchty, Benjamin~F. Jones, and Brian Uzzi.
\newblock The {Increasing} {Dominance} of {Teams} in {Production} of
  {Knowledge}.
\newblock \emph{Science}, 316\penalty0 (5827):\penalty0 1036--1039, May 2007.
\newblock ISSN 0036-8075, 1095-9203.
\newblock \doi{10.1126/science.1136099}.
\newblock URL \url{https://science.sciencemag.org/content/316/5827/1036}.

\bibitem[Jones et~al.(2008)Jones, Wuchty, and
  Uzzi]{jones_multi-university_2008}
Benjamin~F. Jones, Stefan Wuchty, and Brian Uzzi.
\newblock Multi-university research teams: shifting impact, geography, and
  stratification in science.
\newblock \emph{Science (New York, N.Y.)}, 322\penalty0 (5905):\penalty0
  1259--1262, November 2008.
\newblock ISSN 1095-9203.
\newblock \doi{10.1126/science.1158357}.

\bibitem[Gazni et~al.(2012)Gazni, Sugimoto, and Didegah]{gazni_mapping_2012}
Ali Gazni, Cassidy~R. Sugimoto, and Fereshteh Didegah.
\newblock Mapping world scientific collaboration: {Authors}, institutions, and
  countries.
\newblock \emph{Journal of the American Society for Information Science and
  Technology}, 63\penalty0 (2):\penalty0 323--335, 2012.
\newblock ISSN 1532-2890.
\newblock \doi{10.1002/asi.21688}.
\newblock URL \url{https://onlinelibrary.wiley.com/doi/abs/10.1002/asi.21688}.

\bibitem[Ding et~al.(2010)Ding, Levin, Stephan, and Winkler]{ding_impact_2010}
Waverly~W. Ding, Sharon~G. Levin, Paula~E. Stephan, and Anne~E. Winkler.
\newblock The {Impact} of {Information} {Technology} on {Academic}
  {Scientists}' {Productivity} and {Collaboration} {Patterns}.
\newblock \emph{Management Science}, 56\penalty0 (9):\penalty0 1439--1461,
  2010.
\newblock ISSN 0025-1909.
\newblock \doi{10.1287/mnsc.1100.1195}.

\bibitem[Frenken et~al.(2009)Frenken, Hoekman, Kok, Ponds, van Oort, and van
  Vliet]{frenken_death_2009}
Koen Frenken, Jarno Hoekman, Suzanne Kok, Roderik Ponds, Frank van Oort, and
  Joep van Vliet.
\newblock Death of {Distance} in {Science}? {A} {Gravity} {Approach} to
  {Research} {Collaboration}.
\newblock In Andreas Pyka and Andrea Scharnhorst, editors, \emph{Innovation
  {Networks}: {New} {Approaches} in {Modelling} and {Analyzing}}, Understanding
  {Complex} {Systems}, pages 43--57. Springer, Berlin, Heidelberg, 2009.
\newblock ISBN 978-3-540-92267-4.
\newblock \doi{10.1007/978-3-540-92267-4_3}.
\newblock URL \url{https://doi.org/10.1007/978-3-540-92267-4_3}.

\bibitem[Ubfal and Maffioli(2011)]{ubfal_impact_2011}
Diego Ubfal and Alessandro Maffioli.
\newblock The impact of funding on research collaboration: {Evidence} from a
  developing country.
\newblock \emph{Research Policy}, 40\penalty0 (9):\penalty0 1269--1279,
  November 2011.
\newblock ISSN 0048-7333.
\newblock \doi{10.1016/j.respol.2011.05.023}.
\newblock URL
  \url{http://www.sciencedirect.com/science/article/pii/S0048733311001041}.

\bibitem[Kumar(2015)]{kumar_co-authorship_2015}
Sameer Kumar.
\newblock Co-authorship networks: a review of the literature.
\newblock \emph{Aslib Journal of Information Management}, 67\penalty0
  (1):\penalty0 55--73, January 2015.
\newblock ISSN 2050-3806.
\newblock \doi{10.1108/AJIM-09-2014-0116}.
\newblock URL \url{https://doi.org/10.1108/AJIM-09-2014-0116}.

\bibitem[Katz and Martin(1997)]{katz_what_1997}
J.~Sylvan Katz and Ben~R. Martin.
\newblock What is research collaboration?
\newblock \emph{Research Policy}, 26\penalty0 (1):\penalty0 1--18, 1997.
\newblock ISSN 0048-7333.
\newblock \doi{10.1016/S0048-7333(96)00917-1}.

\bibitem[Lee and Bozeman(2005)]{lee_impact_2005}
Sooho Lee and Barry Bozeman.
\newblock The {Impact} of {Research} {Collaboration} on {Scientific}
  {Productivity}.
\newblock \emph{Social Studies of Science}, 35\penalty0 (5):\penalty0 673--702,
  2005.
\newblock ISSN 0306-3127.
\newblock \doi{10.1177/0306312705052359}.

\bibitem[Sooryamoorthy(2017)]{sooryamoorthy_types_2017}
Radhamany Sooryamoorthy.
\newblock Do types of collaboration change citation? {A} scientometric analysis
  of social science publications in {South} {Africa}.
\newblock \emph{Scientometrics}, 111\penalty0 (1):\penalty0 379--400, 2017.
\newblock ISSN 0138-9130.
\newblock \doi{10.1007/s11192-017-2265-6}.

\bibitem[Gazni and Didegah(2011)]{gazni_investigating_2011}
Ali Gazni and Fereshteh Didegah.
\newblock Investigating different types of research collaboration and citation
  impact: a case study of {Harvard} {University}'s publications.
\newblock \emph{Scientometrics}, 87\penalty0 (2):\penalty0 251--265, 2011.
\newblock ISSN 0138-9130.
\newblock \doi{10.1007/s11192-011-0343-8}.

\bibitem[Frenken et~al.(2005)Frenken, H{\"o}lzl, and
  Vor]{frenken_citation_2005}
Koen Frenken, Werner H{\"o}lzl, and Friso~de Vor.
\newblock The citation impact of research collaborations: the case of
  {European} biotechnology and applied microbiology (1988--2002).
\newblock \emph{Journal of Engineering and Technology Management}, 22\penalty0
  (1-2):\penalty0 9--30, 2005.
\newblock ISSN 0923-4748.
\newblock \doi{10.1016/j.jengtecman.2004.11.002}.

\bibitem[O'Clery et~al.(2019{\natexlab{b}})O'Clery, Flaherty, and
  Kinsella]{oclery_modular_2019}
Neave O'Clery, Eoin Flaherty, and Stephen Kinsella.
\newblock Modular structure in labour networks reveals skill basins.
\newblock \emph{arXiv:1909.03379 [econ, q-fin]}, September 2019{\natexlab{b}}.
\newblock URL \url{http://arxiv.org/abs/1909.03379}.
\newblock arXiv: 1909.03379.

\bibitem[Burton and Kebler(1960)]{burton_half-life_1960}
R.~E Burton and R.~W Kebler.
\newblock The ``half-life'' of some scientific and technical literatures.
\newblock \emph{Amer. Doc. American Documentation}, 11\penalty0 (1):\penalty0
  18--22, 1960.
\newblock ISSN 0096-946X.
\newblock OCLC: 4647127545.

\bibitem[Garfield and Sher(1963)]{garfield_new_1963}
E.~Garfield and I.~H. Sher.
\newblock New factors in the evaluation of scientific literature through
  citation indexing.
\newblock \emph{American Documentation}, 14\penalty0 (3):\penalty0 195--201,
  1963.
\newblock ISSN 0096-946X.
\newblock \doi{10.1002/asi.5090140304}.

\bibitem[Kessler et~al.(1962)Kessler, Heart, {National Science Foundation
  (U.S.)}, {Massachusetts Institute of Technology}, {Libraries}, and {Lincoln
  Laboratory}]{kessler_analysis_1962}
Maxwell~Mirton Kessler, F.~E Heart, {National Science Foundation (U.S.)},
  {Massachusetts Institute of Technology}, {Libraries}, and {Lincoln
  Laboratory}.
\newblock \emph{Analysis of bibliographic sources in the physical review (vol.
  77, 1950 to vol. 112, 1958)}.
\newblock Massachusetts Institute of Technology, Cambridge, Mass., 1962.
\newblock OCLC: 339864.

\bibitem[Osgood and Xhignesse(1963)]{osgood_characteristics_1963}
Charles~E. Osgood and Louis~V. Xhignesse.
\newblock \emph{Characteristics of bibliographical coverage in psychological
  journals published in 1950 and 1960}.
\newblock Institute of Communicatios Research, University of Illinois, 1963.

\bibitem[Price(1963)]{price_little_1963}
Derek J. de~Solla Price.
\newblock \emph{Little science, big science}.
\newblock Number 1962 in George {B}. {Pegram} lecture series. Columbia
  University Press, New York, 1963.
\newblock ISBN 978-0-231-08562-5.

\bibitem[Tukey(1962)]{tukey_keeping_1962}
John~W. Tukey.
\newblock Keeping {Research} in {Contact} with the {Literature}: {Citation}
  {Indices} and {Beyond}.
\newblock \emph{Journal of Chemical Documentation}, 2\penalty0 (1):\penalty0
  34--37, 1962.
\newblock ISSN 0021-9576.
\newblock \doi{10.1021/c160004a011}.

\bibitem[Price(1965)]{price_networks_1965}
D.~J. Price.
\newblock {NETWORKS} {OF} {SCIENTIFIC} {PAPERS}.
\newblock \emph{Science (New York, N.Y.)}, 149\penalty0 (3683):\penalty0
  510--515, 1965.
\newblock ISSN 0036-8075.
\newblock \doi{10.1126/science.149.3683.510}.

\bibitem[Jaffe and Trajtenberg(1998)]{jaffe_international_1998}
Adam~B Jaffe and Manuel Trajtenberg.
\newblock International {Knowledge} {Flows}: {Evidence} from {Patent}
  {Citations}.
\newblock Working {Paper} 6507, National Bureau of Economic Research, April
  1998.
\newblock URL \url{http://www.nber.org/papers/w6507}.

\bibitem[Melin and Persson(1996)]{melin_studying_1996}
G.~Melin and O.~Persson.
\newblock Studying research collaboration using co-authorships.
\newblock \emph{Scientometrics}, 36\penalty0 (3):\penalty0 363--377, 1996.
\newblock ISSN 0138-9130.
\newblock \doi{10.1007/BF02129600}.

\bibitem[Gl{\"a}nzel and Schubert(2005)]{glanzel_analysing_2005}
Wolfgang Gl{\"a}nzel and Andr{\'a}s Schubert.
\newblock Analysing {Scientific} {Networks} {Through} {Co}-{Authorship}.
\newblock In Henk~F. Moed, Wolfgang Gl{\"a}nzel, and Ulrich Schmoch, editors,
  \emph{Handbook of {Quantitative} {Science} and {Technology} {Research}: {The}
  {Use} of {Publication} and {Patent} {Statistics} in {Studies} of {S}\&{T}
  {Systems}}, pages 257--276. Springer Netherlands, Dordrecht, 2005.
\newblock ISBN 978-1-4020-2755-0.
\newblock \doi{10.1007/1-4020-2755-9_12}.
\newblock URL \url{https://doi.org/10.1007/1-4020-2755-9_12}.

\bibitem[Heinze and Kuhlmann(2008)]{heinze_across_2008}
Thomas Heinze and Stefan Kuhlmann.
\newblock Across institutional boundaries? {Research} collaboration in {German}
  public sector nanoscience.
\newblock \emph{Research Policy}, 37\penalty0 (5):\penalty0 888--899, 2008.
\newblock ISSN 0048-7333.
\newblock \doi{10.1016/j.respol.2008.01.009}.

\bibitem[Newman(2004)]{newman_coauthorship_2004}
M.~Newman.
\newblock Coauthorship networks and patterns of scientific collaboration.
\newblock \emph{Proceedings of the National Academy of Sciences}, 101\penalty0
  (suppl 1):\penalty0 5200--5205, April 2004.
\newblock ISSN 0027-8424, 1091-6490.
\newblock \doi{10.1073/pnas.0307545100}.
\newblock URL \url{https://www.pnas.org/content/101/suppl_1/5200}.

\bibitem[Fagan et~al.(2018)Fagan, Eddens, Dolly, Vanderford, Weiss, and
  Levens]{fagan_assessing_2018}
Jesse Fagan, Katherine~S. Eddens, Jennifer Dolly, Nathan~L. Vanderford, Heidi
  Weiss, and Justin~S. Levens.
\newblock Assessing {Research} {Collaboration} through {Co}-{Authorship}
  {Network} {Analysis}.
\newblock \emph{Journal of Research Administration}, 49\penalty0 (1):\penalty0
  76--99, 2018.
\newblock ISSN 1539-1590.
\newblock URL \url{http://eric.ed.gov/ERICWebPortal/detail?accno=EJ1181982}.

\bibitem[de~Rassenfosse and Seliger(2020)]{de_rassenfosse_sources_2020}
Ga{\'e}tan de~Rassenfosse and Florian Seliger.
\newblock Sources of knowledge flow between developed and developing nations.
\newblock \emph{Science and Public Policy}, 47\penalty0 (1):\penalty0 16--30,
  February 2020.
\newblock ISSN 0302-3427.
\newblock \doi{10.1093/scipol/scz042}.
\newblock URL \url{https://academic.oup.com/spp/article/47/1/16/5580327}.

\bibitem[Newman(2001{\natexlab{a}})]{newman_scientific_2001}
M.~Newman.
\newblock Scientific collaboration networks. {I}. {Network} construction and
  fundamental results.
\newblock \emph{Physical review. E, Statistical, nonlinear, and soft matter
  physics}, 64\penalty0 (1 Pt 2):\penalty0 016131, 2001{\natexlab{a}}.
\newblock ISSN 1539-3755.

\bibitem[Newman(2001{\natexlab{b}})]{newman_scientific_2001-1}
M.~Newman.
\newblock Scientific collaboration networks. {II}. {Shortest} paths, weighted
  networks, and centrality.
\newblock \emph{Physical Review E}, 64\penalty0 (1):\penalty0 7,
  2001{\natexlab{b}}.
\newblock ISSN 1063-651X.
\newblock \doi{10.1103/PhysRevE.64.016132}.

\bibitem[Newman(2001{\natexlab{c}})]{newman_structure_2001}
M.~Newman.
\newblock The structure of scientific collaboration networks.
\newblock \emph{Proceedings of the National Academy of Sciences}, 98\penalty0
  (2):\penalty0 404--409, January 2001{\natexlab{c}}.
\newblock ISSN 0027-8424, 1091-6490.
\newblock \doi{10.1073/pnas.98.2.404}.
\newblock URL \url{https://www.pnas.org/content/98/2/404}.

\bibitem[Barab{\'a}si et~al.(2002)Barab{\'a}si, Jeong, N{\'e}da, Ravasz,
  Schubert, and Vicsek]{barabasi_evolution_2002}
A.~L. Barab{\'a}si, H.~Jeong, Z.~N{\'e}da, E.~Ravasz, A.~Schubert, and
  T.~Vicsek.
\newblock Evolution of the social network of scientific collaborations.
\newblock \emph{Physica A: Statistical Mechanics and its Applications},
  311\penalty0 (3-4):\penalty0 590--614, 2002.
\newblock ISSN 0378-4371.
\newblock \doi{10.1016/S0378-4371(02)00736-7}.

\bibitem[Fatt et~al.(2010)Fatt, Ujum, and Ratnavelu]{fatt_structure_2010}
Choong Fatt, Ephrance Ujum, and Kuru Ratnavelu.
\newblock The structure of collaboration in the {Journal} of {Finance}.
\newblock \emph{Scientometrics}, 85\penalty0 (3):\penalty0 849--860, 2010.
\newblock ISSN 0138-9130.
\newblock \doi{10.1007/s11192-010-0254-0}.

\bibitem[Racherla and Hu(2010)]{racherla_social_2010}
Pradeep Racherla and Clark Hu.
\newblock A social network perspective of tourism research collaborations.
\newblock \emph{Annals of Tourism Research}, 37\penalty0 (4):\penalty0
  1012--1034, 2010.
\newblock ISSN 0160-7383.
\newblock \doi{10.1016/j.annals.2010.03.008}.

\bibitem[Ye et~al.(2012)Ye, Song, and Li]{ye_cross-institutional_2012}
Qiang Ye, Haiyan Song, and Tong Li.
\newblock Cross-institutional collaboration networks in tourism and hospitality
  research.
\newblock \emph{Tourism Management Perspectives}, 2-3:\penalty0 55--64, 2012.
\newblock ISSN 2211-9736.
\newblock \doi{10.1016/j.tmp.2012.03.002}.

\bibitem[Santos and Santos(2016)]{santos_co-authorship_2016}
Jos{\'e} Ant{\'o}nio~C. Santos and Margarida~Cust{\'o}dio Santos.
\newblock Co-authorship networks: {Collaborative} research structures at the
  journal level.
\newblock \emph{Tourism \& Management Studies}, 12\penalty0 (1):\penalty0
  05--13, 2016.
\newblock ISSN 2182-8466.
\newblock \doi{10.18089/tms.2016.12101}.
\newblock URL
  \url{http://www.scielo.mec.pt/scielo.php?script=sci_arttext&pid=S2182-84582016000100001&lng=en&tlng=en}.

\bibitem[Goh et~al.(2003)Goh, Oh, Kahng, and Kim]{goh_betweenness_2003}
K.-I. Goh, E.~Oh, B.~Kahng, and D.~Kim.
\newblock Betweenness centrality correlation in social networks.
\newblock \emph{Physical review. E, Statistical, nonlinear, and soft matter
  physics}, 67\penalty0 (1 Pt 2):\penalty0 017101, 2003.
\newblock ISSN 1539-3755.
\newblock \doi{10.1103/PhysRevE.67.017101}.

\bibitem[Goh et~al.(2002)Goh, Oh, Jeong, Kahng, and
  Kim]{goh_classification_2002}
Kwang-Il Goh, Eulsik Oh, Hawoong Jeong, Byungnam Kahng, and Doochul Kim.
\newblock Classification of scale-free networks.
\newblock \emph{Proceedings of the National Academy of Sciences of the United
  States of America}, 99\penalty0 (20):\penalty0 12583--12588, 2002.
\newblock ISSN 0027-8424.
\newblock \doi{10.1073/pnas.202301299}.
\newblock URL \url{http://www.pnas.org/content/99/20/12583.abstract}.

\bibitem[Yan and Ding(2009)]{yan_applying_2009}
Erjia Yan and Ying Ding.
\newblock Applying centrality measures to impact analysis: {A} coauthorship
  network analysis.
\newblock \emph{Journal of the American Society for Information Science \&
  Technology}, 60\penalty0 (10):\penalty0 2107--2118, October 2009.
\newblock ISSN 15322882.
\newblock \doi{10.1002/asi.21128}.
\newblock URL
  \url{http://search.ebscohost.com/login.aspx?direct=true&db=bth&AN=44295774&site=ehost-live&authtype=ip,uid}.

\bibitem[Hou et~al.(2008)Hou, Kretschmer, and Liu]{hou_structure_2008}
Haiyan Hou, Hildrun Kretschmer, and Zeyuan Liu.
\newblock The structure of scientific collaboration networks in
  {Scientometrics}.
\newblock \emph{Scientometrics}, 75\penalty0 (2):\penalty0 189--202, 2008.
\newblock ISSN 0138-9130.
\newblock \doi{10.1007/s11192-007-1771-3}.

\bibitem[Rodriguez and Pepe(2008)]{rodriguez_relationship_2008}
Marko~A. Rodriguez and Alberto Pepe.
\newblock On the relationship between the structural and socioacademic
  communities of a coauthorship network.
\newblock \emph{Journal of Informetrics}, 2\penalty0 (3):\penalty0 195--201,
  2008.
\newblock ISSN 1751-1577.
\newblock \doi{10.1016/j.joi.2008.04.002}.

\bibitem[Wagner and Leydesdorff(2005{\natexlab{b}})]{wagner_network_2005}
Caroline~S. Wagner and Loet Leydesdorff.
\newblock Network structure, self-organization, and the growth of international
  collaboration in science.
\newblock \emph{Research Policy}, 34\penalty0 (10):\penalty0 1608--1618,
  December 2005{\natexlab{b}}.
\newblock ISSN 0048-7333.
\newblock \doi{10.1016/j.respol.2005.08.002}.
\newblock URL
  \url{http://www.sciencedirect.com/science/article/pii/S0048733305001745}.

\bibitem[Ribeiro et~al.(2018)Ribeiro, Rapini, Silva, and
  Albuquerque]{ribeiro_growth_2018}
Leonardo~Costa Ribeiro, M{\'a}rcia~Siqueira Rapini, Leandro~Alves Silva, and
  Eduardo~Motta Albuquerque.
\newblock Growth patterns of the network of international collaboration in
  science.
\newblock \emph{Scientometrics}, 114\penalty0 (1):\penalty0 159--179, January
  2018.
\newblock ISSN 1588-2861.
\newblock \doi{10.1007/s11192-017-2573-x}.
\newblock URL \url{https://doi.org/10.1007/s11192-017-2573-x}.

\bibitem[Leydesdorff et~al.(2013)Leydesdorff, Wagner, Park, and
  Adams]{leydesdorff_international_2013}
Loet Leydesdorff, Caroline Wagner, Han~Woo Park, and Jonathan Adams.
\newblock International {Collaboration} in {Science}: {The} {Global} {Map} and
  the {Network}.
\newblock \emph{arXiv:1301.0801 [cs]}, January 2013.
\newblock URL \url{http://arxiv.org/abs/1301.0801}.
\newblock arXiv: 1301.0801.

\bibitem[Doria~Arrieta et~al.(2017)Doria~Arrieta, Pammolli, and
  Petersen]{doria_arrieta_quantifying_2017}
Omar~A. Doria~Arrieta, Fabio Pammolli, and Alexander~M. Petersen.
\newblock Quantifying the negative impact of brain drain on the integration of
  {European} science.
\newblock \emph{Science Advances}, 3\penalty0 (4):\penalty0 e1602232, 2017.
\newblock ISSN 2375-2548.
\newblock \doi{10.1126/sciadv.1602232}.

\bibitem[Chinchilla-Rodr{\'\i}guez et~al.(2012)Chinchilla-Rodr{\'\i}guez,
  Benavent-P{\'e}rez, de~Moya-Aneg{\'o}n, and
  Miguel]{chinchilla-rodriguez_international_2012}
Zaida Chinchilla-Rodr{\'\i}guez, Maria Benavent-P{\'e}rez, F{\'e}lix
  de~Moya-Aneg{\'o}n, and Sandra Miguel.
\newblock International collaboration in {Medical} {Research} in {Latin}
  {America} and the {Caribbean} (2003-2007).
\newblock \emph{Journal of the American Society for Information Science and
  Technology}, 63\penalty0 (11):\penalty0 2223--2238, 2012.
\newblock ISSN 1532-2882.
\newblock \doi{10.1002/asi.22669}.

\bibitem[Geuna(2015)]{geuna_global_2015}
Aldo Geuna.
\newblock \emph{Global {Mobility} of {Research} {Scientists}: {The} {Economics}
  of {Who} {Goes} {Where} and {Why}}.
\newblock Academic Press, August 2015.
\newblock ISBN 978-0-12-801681-7.
\newblock Google-Books-ID: l7rjAwAAQBAJ.

\bibitem[Scherngell(2013)]{scherngell_geography_2013}
Thomas Scherngell, editor.
\newblock \emph{The {Geography} of {Networks} and {R}\&{D} {Collaborations}}.
\newblock Advances in {Spatial} {Science}. Springer International Publishing,
  2013.
\newblock ISBN 978-3-319-02698-5.
\newblock \doi{10.1007/978-3-319-02699-2}.
\newblock URL \url{https://www.springer.com/gp/book/9783319026985}.

\bibitem[Gusenbauer(2019)]{gusenbauer_google_2019}
Michael Gusenbauer.
\newblock Google {Scholar} to overshadow them all? {Comparing} the sizes of 12
  academic search engines and bibliographic databases.
\newblock \emph{Scientometrics}, 118\penalty0 (1):\penalty0 177--214, January
  2019.
\newblock ISSN 1588-2861.
\newblock \doi{10.1007/s11192-018-2958-5}.
\newblock URL \url{https://doi.org/10.1007/s11192-018-2958-5}.

\bibitem[Harzing and Alakangas(2016)]{harzing_google_2016}
Anne-Wil Harzing and Satu Alakangas.
\newblock Google {Scholar}, {Scopus} and the {Web} of {Science}: a longitudinal
  and cross-disciplinary comparison.
\newblock \emph{Scientometrics}, 106\penalty0 (2):\penalty0 787--804, February
  2016.
\newblock ISSN 1588-2861.
\newblock \doi{10.1007/s11192-015-1798-9}.
\newblock URL \url{https://doi.org/10.1007/s11192-015-1798-9}.

\bibitem[Gusenbauer and Haddaway(2019)]{gusenbauer_which_2019}
Michael Gusenbauer and Neal~R. Haddaway.
\newblock Which academic search systems are suitable for systematic reviews or
  meta-analyses? {Evaluating} retrieval qualities of {Google} {Scholar},
  {PubMed}, and 26 other resources.
\newblock \emph{Research Synthesis Methods}, n/a\penalty0 (n/a), 2019.
\newblock ISSN 1759-2887.
\newblock \doi{10.1002/jrsm.1378}.
\newblock URL \url{https://onlinelibrary.wiley.com/doi/abs/10.1002/jrsm.1378}.

\bibitem[{Elsevier}(2020)]{elsevier_scopus_2020}
{Elsevier}.
\newblock Scopus {\textbar} {The} largest database of peer-reviewed literature
  {\textbar} {Elsevier}, 2020.
\newblock URL \url{https://www.elsevier.com/en-gb/solutions/scopus}.

\bibitem[Mingers and Meyer(2017)]{mingers_normalizing_2017}
John Mingers and Martin Meyer.
\newblock Normalizing {Google} {Scholar} data for use in research evaluation.
\newblock \emph{Scientometrics}, 112\penalty0 (2):\penalty0 1111--1121, August
  2017.
\newblock ISSN 1588-2861.
\newblock \doi{10.1007/s11192-017-2415-x}.
\newblock URL \url{https://doi.org/10.1007/s11192-017-2415-x}.

\bibitem[Aksnes and Sivertsen(2019)]{aksnes_criteria-based_2019}
Dag~W. Aksnes and Gunnar Sivertsen.
\newblock A {Criteria}-based {Assessment} of the {Coverage} of {Scopus} and
  {Web} of {Science}.
\newblock \emph{Journal of Data and Information Science}, 4\penalty0
  (1):\penalty0 1--21, February 2019.
\newblock \doi{10.2478/jdis-2019-0001}.
\newblock URL
  \url{https://content.sciendo.com/configurable/contentpage/journals$002fjdis$002f4$002f1$002farticle-p1.xml}.

\bibitem[Bennett(2013)]{bennett_english_2013}
Karen Bennett.
\newblock English as a {Lingua} {Franca} in {Academia}.
\newblock \emph{The Interpreter and Translator Trainer}, 7\penalty0
  (2):\penalty0 169--193, September 2013.
\newblock ISSN 1750-399X.
\newblock \doi{10.1080/13556509.2013.10798850}.
\newblock URL \url{https://doi.org/10.1080/13556509.2013.10798850}.
\newblock Publisher: Routledge \_eprint:
  https://doi.org/10.1080/13556509.2013.10798850.

\bibitem[Trahar et~al.(2019)Trahar, Juntrasook, Burford, Kotze, and
  Wildemeersch]{trahar_hovering_2019}
Sheila Trahar, Adisorn Juntrasook, James Burford, Astrid~von Kotze, and Danny
  Wildemeersch.
\newblock Hovering on the periphery? {Decolonising} writing for academic
  journals.
\newblock \emph{Compare: A Journal of Comparative and International Education},
  49\penalty0 (1):\penalty0 149--167, January 2019.
\newblock ISSN 0305-7925.
\newblock \doi{10.1080/03057925.2018.1545817}.
\newblock URL \url{https://doi.org/10.1080/03057925.2018.1545817}.
\newblock Publisher: Routledge \_eprint:
  https://doi.org/10.1080/03057925.2018.1545817.

\bibitem[Gargiulo et~al.(2016)Gargiulo, Caen, Lambiotte, and
  Carletti]{gargiulo_classical_2016}
Floriana Gargiulo, Auguste Caen, Renaud Lambiotte, and Timoteo Carletti.
\newblock The classical origin of modern mathematics.
\newblock \emph{EPJ Data Science}, 5\penalty0 (1):\penalty0 26, 2016.

\bibitem[Smirnov and Smirnov(1939)]{smirnov_estimation_1939}
Nikolai Smirnov and N.~V. Smirnov.
\newblock On the estimation of the discrepancy between empirical curves of
  distribution for two independent samples.
\newblock \emph{Bull. Math. Univ. Moscou}, 1939.

\bibitem[Müllner(2011)]{mullner_modern_2011}
Daniel Müllner.
\newblock Modern hierarchical, agglomerative clustering algorithms.
\newblock \emph{arXiv preprint arXiv:1109.2378}, 2011.

\bibitem[Evans et~al.(2011)Evans, Lambiotte, and
  Panzarasa]{evans_community_2011}
T.~S. Evans, Renaud Lambiotte, and Pietro Panzarasa.
\newblock Community structure and patterns of scientific collaboration in
  business and management.
\newblock \emph{Scientometrics}, 89\penalty0 (1):\penalty0 381--396, 2011.

\bibitem[Kumar et~al.(1986)Kumar, Kumar, and Kapur]{kumar_normalized_1986}
Uma Kumar, Vinod Kumar, and J.~N. Kapur.
\newblock Normalized measures of entropy.
\newblock \emph{International Journal Of General System}, 12\penalty0
  (1):\penalty0 55--69, 1986.

\bibitem[Neffke and Henning(2013)]{neffke_skill_2013}
Frank Neffke and Martin Henning.
\newblock Skill relatedness and firm diversification.
\newblock \emph{Strategic Management Journal}, 34\penalty0 (3):\penalty0
  297--316, 2013.

\bibitem[Balassa(1965)]{balassa1965trade}
Bela Balassa.
\newblock Trade liberalisation and ``revealed'' comparative advantage1.
\newblock \emph{The Manchester School}, 33\penalty0 (2):\penalty0 99--123,
  1965.

\bibitem[Newman and Girvan(2004)]{newman_finding_2004}
Mark~EJ Newman and Michelle Girvan.
\newblock Finding and evaluating community structure in networks.
\newblock \emph{Physical review E}, 69\penalty0 (2):\penalty0 026113, 2004.

\bibitem[Javed et~al.(2018)Javed, Younis, Latif, Qadir, and
  Baig]{javed_community_2018}
Muhammad~Aqib Javed, Muhammad~Shahzad Younis, Siddique Latif, Junaid Qadir, and
  Adeel Baig.
\newblock Community detection in networks: {A} multidisciplinary review.
\newblock \emph{Journal of Network and Computer Applications}, 108:\penalty0
  87--111, 2018.

\bibitem[Delvenne et~al.(2010)Delvenne, Yaliraki, and
  Barahona]{delvenne2010stability}
J-C Delvenne, Sophia~N Yaliraki, and Mauricio Barahona.
\newblock Stability of graph communities across time scales.
\newblock \emph{Proceedings of the National Academy of Sciences}, 107\penalty0
  (29):\penalty0 12755--12760, 2010.

\bibitem[Lambiotte et~al.(2008)Lambiotte, Delvenne, and
  Barahona]{lambiotte2008laplacian}
Renaud Lambiotte, J-C Delvenne, and Mauricio Barahona.
\newblock Laplacian dynamics and multiscale modular structure in networks.
\newblock \emph{arXiv preprint arXiv:0812.1770}, 2008.

\bibitem[Lambiotte et~al.(2011)Lambiotte, Sinatra, Delvenne, Evans, Barahona,
  and Latora]{lambiotte2011flow}
Renaud Lambiotte, Roberta Sinatra, J-C Delvenne, Tim~S Evans, Mauricio
  Barahona, and Vito Latora.
\newblock Flow graphs: Interweaving dynamics and structure.
\newblock \emph{Physical Review E}, 84\penalty0 (1):\penalty0 017102, 2011.

\bibitem[Reichardt and Bornholdt(2006)]{reichardt2006statistical}
J{\"o}rg Reichardt and Stefan Bornholdt.
\newblock Statistical mechanics of community detection.
\newblock \emph{Physical review E}, 74\penalty0 (1):\penalty0 016110, 2006.

\bibitem[Fortunato and Barthelemy(2007)]{fortunato_resolution_2007}
Santo Fortunato and Marc Barthelemy.
\newblock Resolution limit in community detection.
\newblock \emph{Proceedings of the national academy of sciences}, 104\penalty0
  (1):\penalty0 36--41, 2007.

\bibitem[Blondel et~al.(2008)Blondel, Guillaume, Lambiotte, and
  Lefebvre]{blondel_fast_2008}
Vincent~D. Blondel, Jean-Loup Guillaume, Renaud Lambiotte, and Etienne
  Lefebvre.
\newblock Fast unfolding of communities in large networks.
\newblock \emph{Journal of statistical mechanics: theory and experiment},
  2008\penalty0 (10):\penalty0 P10008, 2008.

\bibitem[Traag et~al.(2019)Traag, Waltman, and van Eck]{traag_louvain_2019}
Vincent~A. Traag, Ludo Waltman, and Nees~Jan van Eck.
\newblock From {Louvain} to {Leiden}: guaranteeing well-connected communities.
\newblock \emph{Scientific reports}, 9\penalty0 (1):\penalty0 1--12, 2019.

\bibitem[Pietilänen and Diot(2012)]{pietilanen_dissemination_2012}
Anna-Kaisa Pietilänen and Christophe Diot.
\newblock Dissemination in opportunistic social networks: the role of temporal
  communities.
\newblock In \emph{Proceedings of the thirteenth {ACM} international symposium
  on {Mobile} {Ad} {Hoc} {Networking} and {Computing}}, pages 165--174, 2012.

\bibitem[Witten and Frank(2002)]{witten_data_2002}
Ian~H. Witten and Eibe Frank.
\newblock Data mining: practical machine learning tools and techniques with
  {Java} implementations.
\newblock \emph{Acm Sigmod Record}, 31\penalty0 (1):\penalty0 76--77, 2002.

\bibitem[G{\'o}mez~Prieto et~al.(2019)G{\'o}mez~Prieto, Demblans, and
  Palazuelos~Mart{\'\i}nez]{gomez_prieto_smart_2019}
Javier G{\'o}mez~Prieto, Albane Demblans, and Manuel Palazuelos~Mart{\'\i}nez.
\newblock Smart {Specialisation} in the world, an {EU} policy approach helping
  to discover innovation globally.
\newblock Technical report, Publications Office of the European Union,
  Luxembourg, 2019.
\newblock URL
  \url{https://s3platform.jrc.ec.europa.eu/-/smart-specialisation-in-the-world-an-eu-policy-approach-helping-to-discover-innovation-globally?inheritRedirect=true}.

\bibitem[Delvenne et~al.(2013)Delvenne, Schaub, Yaliraki, and
  Barahona]{delvenne_stability_2013}
Jean-Charles Delvenne, Michael~T. Schaub, Sophia~N. Yaliraki, and Mauricio
  Barahona.
\newblock The stability of a graph partition: {A} dynamics-based framework for
  community detection.
\newblock In \emph{Dynamics {On} and {Of} {Complex} {Networks}, {Volume} 2},
  pages 221--242. Springer, 2013.

\bibitem[Cover and Thomas(1991)]{cover1991information}
Thomas~M Cover and Joy~A Thomas.
\newblock Information theory and statistics.
\newblock \emph{Elements of Information Theory}, 1\penalty0 (1):\penalty0
  279--335, 1991.

\bibitem[Sch{\"u}tze et~al.(2008)Sch{\"u}tze, Manning, and
  Raghavan]{schutze2008introduction}
Hinrich Sch{\"u}tze, Christopher~D Manning, and Prabhakar Raghavan.
\newblock \emph{Introduction to information retrieval}, volume~39.
\newblock Cambridge University Press Cambridge, 2008.

\end{thebibliography}

\clearpage

\clearpage

\appendix
\numberwithin{equation}{section}
\makeatletter 
\newcommand{\section@cntformat}{Appendix \thesection:\ }
\makeatother

\section{Data processing}

As our data covers over five decades, our analysis spans such geopolitical changes as the reunification of East Germany and West Germany in 1990, the collapse of the Soviet Union in 1991, the breakup of Yugoslavia from 1991 to 1992 and the dissolution of Czechoslovakia to the Czech Republic and Slovakia in 1993 as well as smaller transitions such as post-colonial transitions during the 1970s and early 1980s, the independence of Bangladesh from Pakistan in 1971, Palestinian declaration of independence in 1988, the independence of Namibia from South Africa in 1990, unification of North and South Yemen in 1990, and the independence of Eritrea from Ethiopia in 1993, East-Timor from Indonesia in 1999, and South Sudan from Sudan in 2011. Since we are interested in observing international collaboration across the network of countries over time, some of our methods require the network to remain relatively consistent over time and in order to achieve this, we adjust for the larger geopolitical transitions by keeping the Soviet Union, Yugoslavia, Germany, and Czechoslovakia as single nodes throughout the analysis. 
We consider this operationalisation justified by the fact that beyond fulfilling our methodological requirements, the relationship of these larger regions to the rest of the global academia follows rather constant trends (beyond initial disruptions following the political changes), which gives us confidence that the academic institutions continue working in a relatively similar fashion before and after the changes.

\section{Within region strength \label{app:strengths}}

We may define the total \added{collaboration}\deleted{partner} strength within (resp. outside) the region for each country by
%
\begin{equation} \label{eq:contstrength}
    d^{C}_{in}(i,t)=\sum_{j\in J_u} n^{(t)}_{i,j} , \qquad 
    d^{C}_{out}(i,t)=\sum_{j\in J_o} n^{(t)}_{i,j} ,
\end{equation}
then as previously performed with entropy define the proportion of total strength within the region by
\begin{equation}
    d^{C}_{prop}(i,t)=\frac{d^{C}_{in}(i,t)}{
    d^{C}_{in}(i,t)+d^{C}_{out}(i,t)} .
\end{equation}
\deleted{We analogously compute the proportion of collaborations occuring within the region, $d^{C}_{prop}(i,t)$ by replacing $\hat{p}^{(t)}_{i,j}$ in equation~\eqref{eq:contstrength} with $n^{(t)}_{i,j}$, then performing the same subsequent operation.} Now taking the average of \deleted{each}\added{this} measure over each continent, we display results in Figure~\ref{fig:in_out_strengths}. We see a similar picture to those for our entropy measure\deleted{s}, where \deleted{Europe appears to have opened up slightly, while }Africa and Asia in particular have greatly increased their focus on within-region collaboration.

\begin{figure}[t!]
    \centering
    \includegraphics[width=0.9\textwidth]{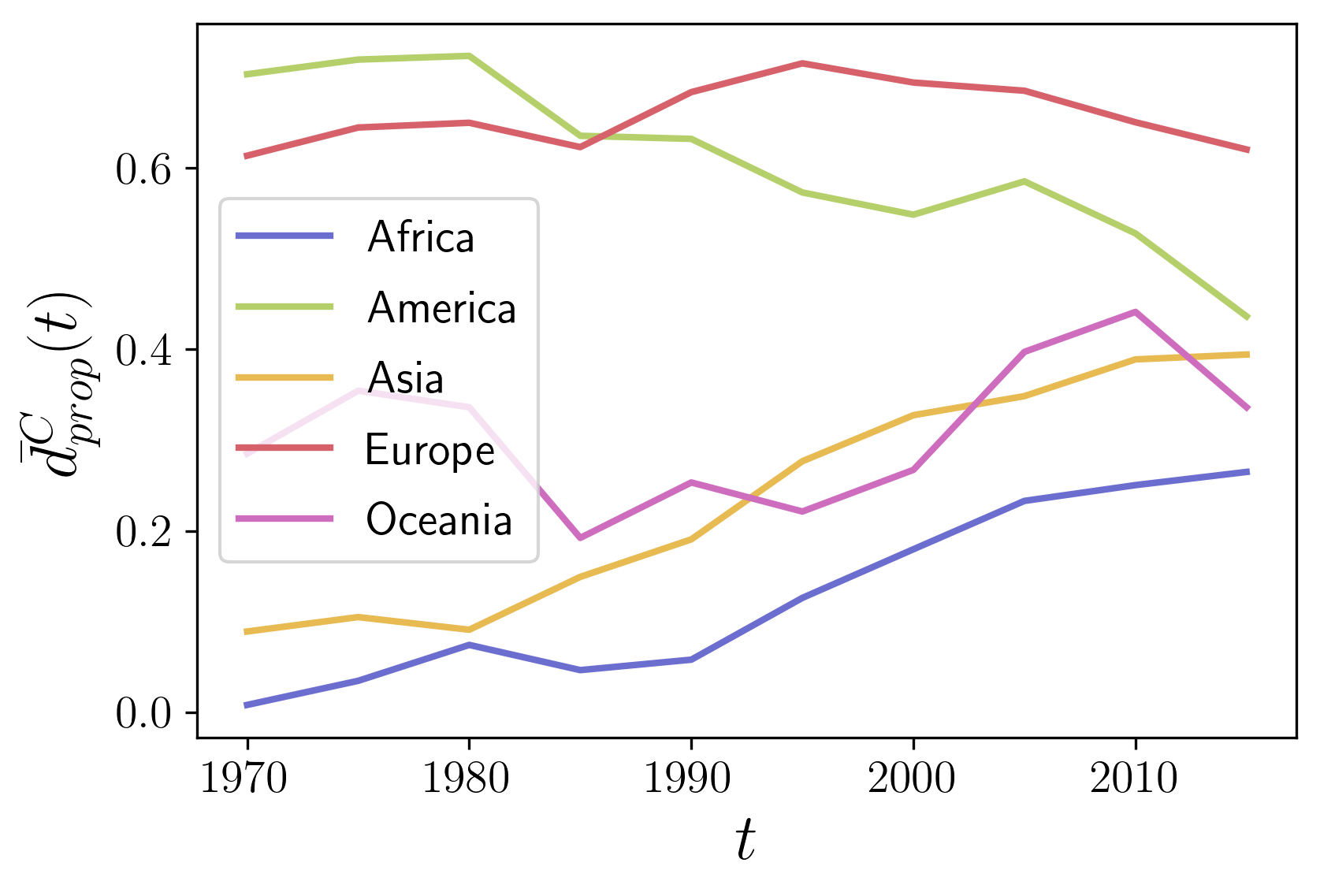}
    \caption{\deleted{Subfigures (a) and (b)}\added{In this figure, we} display the average proportion of within-region strength (as a share of total strength) for each continent\deleted{ for each weight measure}. We observe that Africa and Asia have greatly increased their focus on within-region collaboration since 1970, \added{while countries in America have on average broadened their collaboration profiles}\deleted{particularly as measured by partnership strength}.}
    \label{fig:in_out_strengths}
\end{figure}

\section{Transformation of collaboration counts for visualisation in Figure~\ref{fig:Networks_heatmaps} \label{app:weight_trans}}

\added{For better highlighting significant partnerships when visualising the international collaboration network in Figure~\ref{fig:Networks_heatmaps}, we apply a transformation to the raw counts of collaborations. Specifically, the \emph{collaboration significance} may be defined as}
\begin{equation} 
\label{switches}
s_{i,j}^{(t)}=\frac{n_{i,j}^{(t)}/\sum_{k\neq i}{n_{i,k}^{(t)}}}{\sum_{\ell\neq j}{n_{\ell,j}^{(t)}}/\sum_{m,n}{n_{m,n}^{(t)}}}.
\end{equation}
\added{This corresponds to the ratio of the actual number of collaborations between country pairs to those expected under a configuration model (see \emph{e.g.} }\cite{newman_finding_2004}\added{)---values larger than one correspond to more collaborations occurring than expected at random. As this measure is highly skewed, we re-scale it so that values lie between $-1$ and $1$:}
\begin{equation}
\label{eqn:network}
\hat{p}_{i,j}^{(t)}=\frac{s_{i,j}^{(t)}-1}{s_{i,j}^{(t)}+1}.
\end{equation}
\added{We then finally set $\hat{p}_{i,j}^{(t)}=0$ if $\hat{p}_{i,j}^{(t)}<0$, \emph{i.e.}, those pairs for which fewer collaborations occurred than would be expected, and visualise the resulting network in Figure~\ref{fig:Networks_heatmaps}.}

\section{Using stability in community detection \label{app:stability}}

\added{The key step is the relation of modularity to the stability of communities under Laplacian dynamics} \cite{lambiotte2008laplacian}\added{, as defined by the formula} 
\begin{equation} \label{eq:lapdyn}
  	\dot{p}_{i} = \sum_{j} L_{ij}p_{j} .	
\end{equation}
\added{Here the appropriate matrix $L$ to relate to modularity depends on the type of network under consideration. As our network is undirected, we may use the normalised Laplacian matrix}
\begin{equation}
  	L_{ij} = \frac{A_{ij}}{k_{j}} - \delta_{ij} ,
\end{equation}
\added{and the stability of the partition is then defined to be} \cite{lambiotte2008laplacian}
\begin{equation} \label{lappartstab}
  	R(\tau) = \sum_{i, j} \left[(\e ^{\tau L})_{ij}p_{j}^{*} - p_{i}^{*}p_{j}^{*}\right]\delta(g_{i}, g_{j}) ,
\end{equation}
\added{where $p_{i}^{*} = k_{i}/2m$ is the stationary solution to \eqref{eq:lapdyn}. Expanding the exponential matrix to first order in $\tau$, }
\begin{equation}
  	(\e ^{\tau L})_{ij} = \delta_{ij} + \tau L_{ij} + \Oh(\tau^{2}) ,
\end{equation}
\added{thus naturally leads to the inclusion of a \emph{resolution parameter} $\gamma = 1/\tau$ in the modularity equation, as ignoring the constant term and dividing by $\tau$ we then have}
\begin{equation} \label{eq:lin_stab}
     R_{lin}(\tau)= \frac{1}{2m}\sum_{i, j} \left\{A_{ij} - \gamma\frac{k_{i}k_{j}}{2m} \right\}\delta(g_{i}, g_{j}),
\end{equation}
\added{an answer to the resolution problem previously mentioned. }

\added{In matrix form, equation \eqref{eq:lin_stab} may be written as }\cite{delvenne_stability_2013}
\begin{equation}
     R_{lin}(\tau)=\mathrm{Tr}\left[H^\top \left(\frac{A}{2m} - \gamma\pi^\top\pi\right)H\right], 
\end{equation}
\added{where $H$ is the $N\times c$ matrix with $H_{ij}=1$ if country $i$ is in community $j$ and zero else, and $\mathrm{Tr}$ corresponds to the trace operator. For a given Markov time $\tau$ (or equivalently resolution parameter $\gamma$), we seek the partition that maximises this function. As this optimisation problem is NP-hard, as stated in the main text, we use a greedy method from }\cite{traag_louvain_2019}.

\added{Suitable resolution parameter ranges are typically discovered through calculation of the variation of information. For two partitions $X$ and $Y$ of a set $A$ into disjoint subsets, $X = \{ X_1, X_2, \dots, X_k\}$ and $Y=\{Y_{1},Y_{2},..,,Y_{l}\}$, this measure is defined as follows. Let $n=\sum _{i}|X_{i}|=\sum _{j}|Y_{j}|=|A|$, $ p_{i}=|X_{i}|/n$, $q_{j}=|Y_{j}|/n$, $r_{ij}=|X_{i}\cap Y_{j}|/n$. Then the normalised variation of information between the two partitions is:}
\begin{equation}
     \mathrm {VI} (X,Y)=-\frac{1}{\log N}\sum _{i,j}r_{ij}\left[\log(r_{ij}/p_{i})+\log(r_{ij}/q_{j})\right].
\end{equation}
\added{As there is some stochasticity in the output of the optimisation process, we run the method many times for each resolution, and collect the resulting partitions. If the average variation of information between each pair of such partitions is small, then this suggests that this parameter choice provides somewhat more robust communities.}

\section{How good a null model is the configuration model? \label{app:kl-div}}

\added{As suggested in the main text, the grouping of major producers -- specifically the USA and China -- together in a single community in recent years, despite not necessarily having similar partners other than each other, may imply that the configuration null model used is not the optimal choice of null model for uncovering significant partnerships. To further investigate this, we may study how closely the observed distribution of collaborations for each country lies to that predicted by the configuration model. One way of assessing the proximity of two such probability distributions is the Kullback-Liebler (KL) divergence (see \emph{e.g.} }\cite{cover1991information}\added{). For two discrete probability distributions $P$ and $Q$ that have the same support, $\chi$ say, to find the information gain from using $P$ (which can be the real observed data) in place of our model, $Q$, we calculate}
\begin{equation} \label{eq:kl-div}
	D_{\text{KL}}(P \ \Vert \ Q) = \sum_{x\in\chi} P(x)\log\left(\frac{P(x)}{Q(x)}\right) .
\end{equation}
\added{In our situation, for country $i$ in year $t$, we compare the empirical distribution of collaborations with all other countries, \emph{i.e.} $P_{i}^{(t)}(j)=p^{(t)}_{ij}=n^{(t)}_{ij}/\sum_{k}n^{(t)}_{ik}$, to that predicted by the configuration null model, where the expected number of links between countries $i$ and $j$ is $k_{i}^{(t)}k_{j}^{(t)}/2m^{(t)}$ (followed by analogous normalisation for each country to form a probability distribution). To ensure the support of the empirical distribution and the configuration model match, we perform additive smoothing (see \emph{e.g.} }\cite{schutze2008introduction})\added{, \emph{i.e.} we add one to the count of collaborations between each pair of countries prior to normalising.}

\added{In Figure~\ref{fig:kl-div} we display the average of the resulting KL divergence for each continent across time. We observe that while in 1970 the configuration model was a comparably good choice for all continents, there has since been a large deviation. Europe has on average increasingly collaborated as the model would predict, suggesting that preferential attachment is a major mechanism driving international collaborations at the aggregate level, while other continents -- particularly Africa -- have collaborated in more and more `surprising' patterns relative to the model. This decreasingly suitable description of some regions true collaboration implies that an alternative could be used to further highlight groups of closely partnered countries, though in doing so note we would lose the dynamical interpretation of communities, and the associated stability function. We leave the development of a suitable alternative model to future work.}

\begin{figure}[t!]
    \centering
    \includegraphics[width=0.9\textwidth]{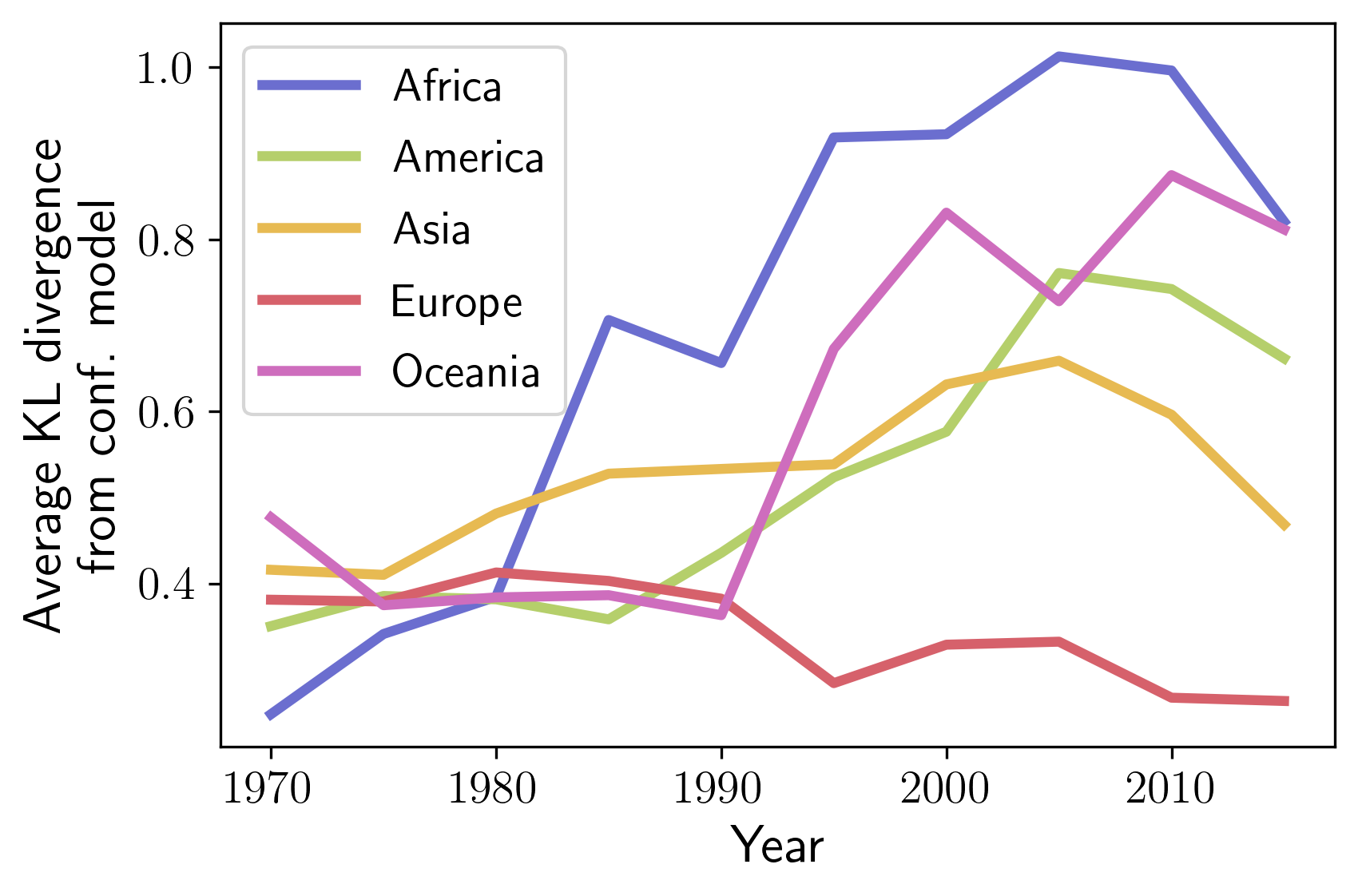}
    \caption{Average KL divergence (as defined in equation~\eqref{eq:kl-div}) for each continent between the observed collaboration counts and those expected by the configuration model. Over time, the configuration model has become a better description for Europe, while other continents have further differed.}
    \label{fig:kl-div}
\end{figure}

\end{backmatter}
\end{document}